\documentstyle[aps]{revtex}

\begin{document}



\title{ Transition from hadronic to partonic interactions for a
composite spin-1/2 model of a nucleon }


\author{J. A. Tjon}
\address{Institute of Theoretical Physics
University of Utrecht, TA 3508 Utrecht, The Netherlands 
and KVI, University of Groningen, 9747 AA Groningen, The Netherlands} 
\author{S. J. Wallace }
\address{Thomas Jefferson National Accelerator Facility, 
12000 Jefferson Ave., Newport News, VA 23606 and Department of Physics
and Center for Theoretical Physics,
University of Maryland, College Park, MD 20742} 

\date{February 17, 2000}
\maketitle
\vspace{-2.3in}

\hspace{3.5in} {\bf JLAB-THY-00-06,~~UMD PP 00-056}
\vspace{2.2in}

\begin{abstract}
A simple model of a composite nucleon is developed in which a
fermion and a boson, representing quark and diquark constituents
of the nucleon, form a bound state owing to a contact interaction.
Photon and pion couplings to the quark provide vertex functions
for the photon and pion interactions with the composite nucleon. 
By a suitable choice of cutoff parameters of the model, realistic 
electromagnetic form factors are obtained.  When a pseudoscalar pion-quark
coupling is used, the pion-nucleon coupling is predominantly pseudovector.
A virtual photopion amplitude is considered in which there are two
types of contributions: hadronic contributions where the photon and pion 
interactions have an intervening propagator of the nucleon or its excited 
states, and contact-like contributions where the photon and pion 
interactions occur within a single vertex.  At large Q, the contact-like 
contributions are dominant.  The model nucleon exhibits scaling behavior
in deep-inelastic scattering and the normalization of the parton
distribution provides a rough normalization of the contact-like
contributions.  Calculations for the virtual photopion amplitude 
are performed using kinematics appropriate to its occurrence as a
meson-exchange current in electron-deuteron scattering.
The results show that the contact-like terms can dominate the 
meson-exchange current for Q $>$ 1 GeV/c.  There is a direct connection of
the contact-like terms to the off-forward parton distributions 
of the model nucleon.  
\end{abstract}




\section{Introduction}

At low energies and momentum transfers, nuclei are described in term of 
nucleons~\cite{Pudlineretal97,CarlsonSchiavilla98}. 
Interactions between the nucleons are modelled 
successfully by  exchange of mesons~\cite{ft,Nijmegen,Machleidt89},
or more simply, by potentials.  
When nuclei are probed at very high momentum transfer, e.g., 
in electron scattering, partons within the nucleons and mesons become the 
dominant scatterers~\cite{Bjorken69,Feynman72}.
Interactions between the partons are described by QCD.  
Between the high and low momentum transfer regimes, there is
a transition region where a good description is lacking.  
The meson-exchange dynamics does not account in a 
satisfactory way for the compositeness of the nucleons and mesons.
Therefore, it is of interest to study quark-based composite models 
of hadrons in order to get some insight on the limits of validity 
of a hadronic description.
    Electron scattering data for momentum transfer 
Q $\approx$ 1 GeV/c often meet dual descriptions:
models based on hadrons on one hand and models based on quark 
phenomenology on the other~\cite{D-photodisintegration98}.  
Moreover, the two kinds of description generally are not 
reconciled to one another in the sense that 
there is no smooth transition from one 
to the other as Q increases.  Perturbative QCD 
descriptions are mainly qualitative 
and not properly normalized at low energy~\cite{IsgurLSmith}.
In the mesonic description,
the mechanism of hard scattering from quarks that predominates in the 
perturbative QCD description is hidden or absent.

In this paper, we develop a simple model of a nucleon as a 
bound state of a fermion and a boson with 
the goal of gaining some insight into the transition region where, 
as Q increases, one 
passes from the dominance of hadronic processes to the 
dominance of scattering from the 
constituents of a nucleon.  One may think of this 
model as having a quark and a spin-0 
diquark bound together to make a nucleon and its excited 
states.  The model is 	covariant and gauge invariant,
but it lacks confinement.  Excited states of the 
nucleon are a continuum of quark and diquark scattering 
states.  Thus, it is mainly useful for processes
where nucleon resonances do not play an important role.
One such case is mesonic-exchange currents in nuclei.  

  An essential feature arises from compositeness: 
there are contact-like terms in second-order interactions.
These are required by gauge invariance and they play 
a small but significant role at low energy, 
for example, in low-energy theorems~\cite{Low58,WGT,PhillipsBirseWallace}.
For very large momentum
transfer, the contact-like terms become dominant.  They contain 
the leading-order mechanism for the external probe to 
scatter from the partons without any 
intervening hadronic state.  When a hadronic state exists between  
interactions, it produces form factors that 
fall rapidly with increasing Q, 
thus quenching the scattering.  This is the fate of hadron-like
terms in the second-order interactions, i.e., 
the terms that provide a hadronic 
interpretation at low momentum transfer.

 In the limit that one of the interactions 
transfers a large momentum, the contact-like terms tend to the off-forward 
parton distributions for the composite nucleon model~\cite{XJi97,Radyushkin96}.
For the simple model that we consider, there is a clean  
separation of the hadron-like and contact-like contributions to 
second-order interactions.  
   Interactions of the model nucleon with an electromagnetic 
probe have some realistic features.  By introducing
cutoff parameters, the nucleon's charge and magnetic 
form factors can be described reasonably.  At low momentum transfers,
interactions of the model nucleon can be interpreted 
in terms of hadron dynamics.  For asymptotically 
large momentum transfer Q, at fixed $x = Q^2/(2 M \nu)$, scaling obtains. 
We calculate the resulting parton distribution $f(x)$.   

In Sec. 2 we formulate the model in terms of a lagrangian for a 
fermion and a boson
interacting via a contact interaction.   The model is 
not renormalizable: it is regulated 
by introducing subtraction terms of the Pauli-Villars 
type.\cite{PauliVillars}  We consider only the simplest 
subset of contributions to the fermion-boson correlator. 
This produces a spin-1/2 
propagator with a single bound state pole (``the nucleon'') 
at mass M.  Electromagnetic and 
pionic interactions are introduced in Sec. 3 as couplings to the 
fermion constituent (``the quark'').
For simplicity, couplings to the boson (``the diquark'') are omitted.  For  
pseudoscalar coupling of the pion to the quark, the model produces mostly 
pseudovector coupling to the nucleon.  It would be purely 
pseudovector if the mass of the quark were zero and there were no
regulators of fermion type.  

    In Sec. 4 we consider a virtual photopion amplitude involving 
second-order interactions with the composite nucleon.  Two types 
of interaction occur: first, interactions with  
intervening propagation of a nucleon or its 
excited states and second, contact-like 
contributions where photon and pion interactions 
with the nucleon occur within the same vertex.  
A standard analysis based upon elementary particles with form factors is
compared with the composite nucleon analysis.  In Sec. 5 we 
consider deep inelastic scattering from the 
nucleon for finite Q and as Q $\to \infty$.   
In Sec. 6 we present calculations of the virtual photo-pion 
production amplitude for a kinematical situation
that arises in meson-exchange contributions 
to electron-deuteron scattering.   
Calculations show that contact-like terms can become 
dominant for Q $\approx$ 1 GeV/c for some 
processes.   Conclusions are presented in 
Sec. 7.  A more complete description of the
details of the calculations is given in four appendices.

\section{Composite nucleon model}

A fermion and a boson interacting via a contact interaction can generate a 
composite spin-1/2 particle.
For this purpose, the following lagrangian is used \cite{WGT}.
\begin{eqnarray}
L = \bar{\psi}(x) ( i \partial \!\!\!/ ~- m) \psi(x) 
+ {1 \over 2} [ \partial_{\nu}\phi (x)
\partial^{\nu}\phi (x)
 - \mu^2 \phi^2(x) ] + g \bar{\psi} (x) \psi(x) \phi^2(x),
\end{eqnarray}
where $\psi(x)$ is the field for a fermion of mass m and $\phi(x)$ is 
the field for a boson of mass $\mu$.  The fermion-boson 
contact interaction with coupling constant g is 
not renormalizable;  finite results are 
obtained by introducing a Pauli-Villars regulator of mass $\Lambda_1$.

A bound state appears as a pole in the fermion-boson correlator,
\begin{equation}
G(p) = i\int d^4x e^{-i p \cdot x} \langle 0 | 
T\left( \psi(x) \phi(x) \bar{\psi}(0) \phi(0)
\right) | 0 \rangle.
\end{equation}
Figure~\ref{fig:green} shows the sequence of elementary 
bubble graphs that contribute to G(p) in a perturbative expansion. Because
this sequence is sufficient to exhibit a bound state, contributions
beyond those shown in Fig. 1 are not considered.

Summing the bubble graphs of Fig. \ref{fig:green} produces
\begin{equation}
G(p) = \frac{1}{1 - \Sigma (p)}.
\end{equation}
Here, $\Sigma (p)$ is the contribution of a single fermion-boson loop, 
\begin{eqnarray}
\Sigma_b (p; m, \mu, \Lambda_1)  = 
          i g \int \frac{d^4k}{(2 \pi)^4} S(p-k; m) D(k; \mu, \Lambda_1),
\label{eq:SigmaBubble}
\end{eqnarray}
where the propagator for the fermion is 
$S(p;m) = 1 /( p\!\!\!/ ~- m + i \eta)$.   
With a Pauli-Villars regulator of mass $\Lambda_1$ 
included, the propagator for the boson line is 
\begin{equation}
D(k;\mu, \Lambda_1) = \frac{1}{k^2 - \mu^2 + i \eta} -  
\frac{1}{k^2 - \Lambda_1^2 + i \eta}.
\end{equation}

A generalization of the model that is suitable for 
describing a nucleon's form factor is 
obtained by including additional regulator terms as follows,
\begin{eqnarray}
\Sigma (p) = i g \int \frac{d^4k}{(2 \pi)^4} 
\Big[ S(p-k; m) - \alpha S(p-k; m_1) 
- (1 - \alpha) S(p-k;m_2) \Big]
\nonumber \\ \times 
\Big[ D(k; \mu, \Lambda_1) + \beta D(k; \Lambda_1,\Lambda_2) \Big].
\nonumber \\ && 
\label{eq:SigmaGen}
\end{eqnarray}
where 
\begin{equation}
\alpha = \frac{m_2 - m}{m_2 - m_1} ,
\end{equation}
and \begin{equation}
\beta = \frac{\mu^2 -  \Lambda_1^2}{\Lambda_2^2 - \Lambda_1^2}.
\end{equation}
The constants $\alpha$ and $\beta$ are selected so that 
high loop momentum is cut off 
as k$^{-9}$.  It is evident that the generalized 
form for $\Sigma (p)$ is equal to a 
linear combination of the elementary bubble graph 
terms, $\Sigma_b$, defined above.
\begin{eqnarray}
&&\Sigma (p) =~~~ \Sigma_b (p;m,\mu, \Lambda_1) 
+~ \beta \Sigma_b(p;m,\Lambda_1,\Lambda_2) \nonumber \\
&&~~~~~~-\alpha \Big[ \Sigma_b (p;m_1,\mu, \Lambda_1) 
+ \beta \Sigma_b(p;m_1,\Lambda_1,\Lambda_2) \Big] \nonumber \\
&&-(1 - \alpha)\Big[ \Sigma_b (p;m_2,\mu, \Lambda_1) 
+ \beta \Sigma_b(p;m_2,\Lambda_1,\Lambda_2) \Big]
\label{eq:Sigmabsum}
\end{eqnarray}

   When a bound state of mass M is present, the pole 
in the composite system propagator G(p) has the form
\begin{equation}
G(p) = \frac{Z_2}{p\!\!\!/ ~- M + i \eta} + R(p),
\label{eq:Gpole}
\end{equation}
where Z$_2$ is a wave-function renormalization factor.  A 
renormalized propagator $\widetilde{ \rm G}$(p) is 
obtained by dividing G(p) by Z$_2$ such that there is unit 
residue for the nucleon pole.  
The remainder R(p) is regular at 
$p\!\!\!/~ = M$ and it represents excited state contributions.  In the model 
considered, the excited state spectrum is a 
continuum of quark-diquark scattering states.  
This is an unrealistic feature for a nucleon so the model should be used  
where the effects of resonances are not important. 

   The most general form for $\Sigma$ 
that is allowed by Lorentz invariance is
\begin{equation}
\Sigma (p) = A(p^2)p\!\!\!/ ~+ B(p^2).
\label{eq:SigmaAB}
\end{equation}
  Presence of the bound state pole means that 
$\Sigma (p) = 1$ at $p\!\!\!/~ = M$.  This condition leads to 
\begin{eqnarray}
Z_2^{-1}  &=& \left( - \frac{d \Sigma (p) }{d p\!\!\!/ }
\right)_{p\!\!\!/~ = M} \nonumber \\
  &=& -[A_0 + 2M(A'_0 M + B'_0 ) ] ,
\label{eq:Zsub2} 
\end{eqnarray}
where $A_0 \equiv A(M^2)$, $A'_0 \equiv dA(p^2)/dp^2|_{p^2 = M^2}$, 
and similarly for $B'_0$.

For later use, we introduce covariant projection operators,
\begin{equation}
L^{\rho} (p) \equiv \frac{ W_p + \rho p\!\!\!/ }{2W_p},
\end{equation}
where $\rho$ = $+$ or $-$, W$_p$ = $\sqrt{{\rm p}^2}$, 
and L$^+$ (p) + L$^-$(p) = 1.  
Projecting the propagator to the $\rho$ = $+$ and $-$ subspaces 
in which $p\!\!\!/$ takes the values $\pm$W$_p$, 
leads, for the renormalized propagator, to
\begin{equation}
 \widetilde{G}(p) = G^+(p) + G^-(p) 
\end{equation}
where
\begin{equation}
G^{\rho}(p) = \frac{L^{\rho}(p)}{Z_2 [ 1 - B(p^2) - \rho W_p A(p^2) ]}.
\end{equation}

To summarize this section, the composite model of a nucleon 
is formulated in a covariant way as a bound state of a spin-1/2 quark and a
spin-0 diquark.  Details of the calculation of A(p$^2$), B(p$^2$) 
and Z$_2$ are given in Appendix A.

\section{Photon and pion interactions}

Electromagnetic interactions are introduced via a 
fermion-photon coupling term
in the lagrangian: $ {\cal L}_{\mu} \equiv$~$\bar{\psi}$(x) 
$\gamma^{\mu} \hat{e}$A$_{\mu}$(x) $\psi$(x), where $\hat{e} = 
{1 \over 2}e(1 + \tau_3)$ is the charge operator for the quark. 
Photon coupling to the boson is omitted in order to keep the model simple.
Consequently, the model proton is composed of a quark of charge e 
and a neutral diquark.  The model neutron is composed of a neutral quark and diquark
and thus has no electromagnetic interations. 

Inserting a photon into the propagator as indicated 
in Figure \ref{fig:phvertex} produces the form
\begin{equation}
G(p_f) \hat{e} \Lambda_{\mu} (p_f,p_i) G(p_i),
\end{equation}
where $\Lambda_{\mu}$ describes the photon-nucleon vertex. 
One extracts the photon-nucleon (dressed) interaction  
as the residue of the two poles 
at $p\!\!\!/_i$ = M and $p\!\!\!/_f$ = M, which leads to
\begin{equation}
\bar{u}(p_f;M) \hat{e} Z_2 \Lambda_{\mu} (p_f,p_i) u(p_i;M),
\end{equation}
where u(p;M) is the Dirac spinor for mass M and momenta 
p$_i$ and p$_f$ are on 
the mass shell.   The Z$_2$ 
factor and Dirac spinor factors arise from the 
parts of the initial- and final-state 
propagators that attach to the vertex $\Lambda_{\mu}$.  
It is convenient to absorb 
the Z$_2$ factor into $\Lambda_{\mu}$ to obtain 
a renormalized vertex $\widetilde{\Lambda}_{\mu}$.
For momenta p$_i$ and 
p$_f$ that are either 
on-shell or off-shell, the renormalized vertex 
involves a fermion-boson loop with a photon 
insertion in the fermion propagator as follows,
\begin{eqnarray}
\widetilde{\Lambda}_{\mu}(p_f,p_i) =&&  i g Z_2 
\int \frac{d^4k}{(2\pi)^4} S(p_f-k;m) \gamma_{\mu} 
S(p_i-k;m) 
D(k;\mu,\Lambda).
\label{eq:LambdamuLoop} 
\end{eqnarray}
In the generalized model with additional Pauli-Villars 
regulators, the vertex is a 
sum of such terms, one for each term in Eq.~(\ref{eq:Sigmabsum}).  
In each $\Sigma_b$, the
fermion propagator $S(p-k;m_n)$ for mass m$_n$ 
is replaced by $S(p_f-k;m_n) \gamma_{\mu} S(p_i-k;m_n)$. 
Gauge invariance requires that the vertex satisfy the following 
Ward-Takahashi identity \cite{Ward50,Takahashi57},
\begin{eqnarray}
(p_f - p_i)^{\mu}\Lambda_{\mu}(p_f,p_i) &=&  G^{-1}(p_f) -
 G^{-1}(p_i) \nonumber \\
   &=& \Sigma (p_i) - \Sigma (p_f).
\label{eq:vertex-WTI}
\end{eqnarray}
This is satisfied when the photon couples to all 
fermion propagators in the same way, 
including those introduced as Pauli-Villars regulators.  

In general, the vertex function can be decomposed in 
terms of charge and magnetic form 
factors F$_1$ and F$_2$.  For the off-mass-shell 
case, there is an additional form factor F$_3$.
Moreover, all form factors depend upon $p\!\!\!/_i$ 
and $p\!\!\!/_f$.   In order to have 
scalar form factors, it is necessary to project 
with the operators L$^{\pm}$ and to 
commute the  $p\!\!\!/_i$ and $p\!\!\!/_f$ toward the projectors so that 
they may be replaced by $\rho_i$W$_i$ or $\rho_f$W$_f$, 
where W$_i$ = $\sqrt{p_i^2}$ and  W$_f$ = $\sqrt{p_f^2}$.  This 
analysis is carried out in Appendix B.  It produces 
\begin{equation}
\widetilde{\Lambda }_{\mu}(p_f,p_i) = \sum _{\rho_f, \rho_i = \pm} 
L^{\rho _f} (p_f)
\Lambda^{\rho_f,\rho_i}_{\mu} (p_f,p_i) L^{\rho} (p_i) ,
\end{equation}
where 
\begin{eqnarray}
\Lambda^{\rho_f,\rho_i}_{\mu} (p_f,p_i) &&= \gamma_{\mu} 
F_1^{\rho _f, \rho_i}(p_f,p_i) + i \sigma_{\mu \nu} q^{\nu} 
F_2^{\rho _f, \rho_i}(p_f,p_i)
 + q_{\mu} F_3^{\rho _f, \rho_i}(p_f,p_i), 
\label{eq:Lambdamu}
\end{eqnarray}
and q = p$_i$ - p$_f$.  
Each scalar form factor is a different function depending on the 
values of $\rho_f$ and $\rho_i$, e.g., F$_1^{+-}$
is different from F$_1^{++}$.  We shall return to this point shortly.  

For the on-shell situation, owing to time-reversal invariance, one only has 
F$_1^{++}$(q$^2$) and F$_2^{++}$(q$^2$),
which are the usual charge and magnetic form factors of the proton.  With 
three fermion masses and three boson masses
as parameters, the generalized model allows a  reasonable fit to 
the proton's electromagnetic form factors.  Figure~\ref{fig:FormFacs} shows 
F$_1^{++}$(q$^2$) and 	F$_2^{++}$(q$^2$) in comparison with
the dipole form F$_{\rm dipole}$ = (1 + Q$^2$/0.71 GeV$^2$)$^{-2}$
that often is used to characterize experimental form factors.  The parameter 
values used are: m= .38, m$_1$ = .56, m$_2$ = .61, $\mu$ = .79, 
$\Lambda_1$ = .85 and $\Lambda_2$ = .90, all in GeV.  The bound state is
at M = .93826 GeV.  The anomalous magnetic moment of the composite nucleon 
is $\kappa = 2 M F_2^{++} (0) = 2.086$, which may 
be compared with $\kappa_{\rm proton} = 1.79$.

A parallel analysis may be made for couplings
of an elementary pion to the quark by adding a 
pseudoscalar $\pi$-quark interaction 
${\cal L}_5\equiv $g$_{\pi}\bar{\psi}$(x)$\gamma_5 
\vec{\tau}\psi$(x) $\cdot \vec{\pi}$(x) to the lagrangian.  
Figure~\ref{fig:phvertex} shows one pion insertion into the 
propagator.  This produces 
\begin{equation}
G(p_f) g_{\pi} \vec{\tau}\cdot \hat{\phi} \Lambda_5(p_f,p_i) G(p_i),
\end{equation}
where $\vec{\tau} \cdot \hat{\phi} = \tau_+\phi_- + \tau_
-\phi_+ + \tau_3\phi_0$, with $\phi_{\pm}$ and $\phi_0$ being isospin wave
functions for $\pi^{\pm}$ and $\pi^0$ mesons.  
A renormalized pion-nucleon vertex function is calculated 
from a fermion-boson loop graph with 
a pseudoscalar insertion on the fermion, as follows,
\begin{eqnarray}
\widetilde{\Lambda}_5(p_f,p_i) =&&i g Z_2 \int \frac{d^4k}{(2 \pi)^4} 
S(p_f-k;m) \gamma_5 S(p_i-k;m) 
D(k;\mu, \Lambda_1) .
\label{eq:Lambda5Loop}
\end{eqnarray}
In the generalized model, the pion-nucleon vertex function is a 
sum of such terms, one for each term 
in Eq.~(\ref{eq:Sigmabsum}).  In each $\Sigma_b$, 
the fermion propagator $S(p-k;m_n)$ 
for mass m$_n$ is replaced 
by $S(p_f-k;m_n) \gamma_5 S(p_i-k;m_n)$.

Again it is necessary to rearrange terms and to project in order to have 
scalar form factors.  
This produces (see Appendix B for details) 
\begin{equation}
\widetilde{\Lambda}_5 (p_f,p_i) =\sum_{\rho_f, \rho_i = \pm} 
L^{\rho_f} (p_f) \gamma_5 
F_5^{\rho _f,\rho_i} (p_f,p_i) L^{\rho_i}(p_i) .
\end{equation}
Figure~\ref{fig:FormFacs} shows the resulting $\pi$N 
form factor F$_5^{++}$(q$^2$)
for on-mass-shell nucleon
momenta.  It is quite similar to the magnetic form factor.

When one leg of the vertex function is off the 
mass shell, the form factors differ 
from the on-shell results.  We wish to relate
the off-shell effects to those appropriate to
a hadronic vertex that is sandwiched between elementary
Dirac propagators. For this purpose, it 
is necessary to incorporate off-shell effects from the propagators
into the off-shell vertex function and form factors. 

In general, one encounters an off-shell vertex function sandwiched 
between propagators, as follows,
\begin{equation}
\widetilde{G}(p_f) \widetilde{\Lambda}(p_f,p_i) \widetilde{G}(p_i). 
\label{eq:G-Lambda-G}
\end{equation}
The renormalized propagator of the composite system may be written as
\begin{equation}
\widetilde{G}(p) = \frac{Z^{(+)} (p) L^{(+)} (p)}{Z_2(W_p - M)} + 
\frac{Z^{(-)} (p) L^{(-)} (p)}{Z_2( -W_p - M)}, 
\end{equation}
where $Z^{(\pm)}(p) = (\pm W_p - M)/[1 - A(p^2) \pm W_p B(p^2)]$ 
are scalar functions.
In the limit that $W_p \rightarrow +M$, $Z^{(+)} 
\rightarrow Z_2$, and in the limit that  $W_p \rightarrow 
-M$, $Z^{(-)} \rightarrow Z_2$.  For a point particle the factors 
$Z^{(\pm)}/Z_2$ are unity, i.e., an 
elementary Dirac propagator may be written in the same 
way with $Z^{(\pm)}/Z_2$ factors replaced by unity.  
Thus, these factors carry off-shell effects
due to the propagator.   A factor $\sqrt{Z^{(\pm)}(p)/Z_2}$ from
each propagator in Eq.~(\ref{eq:G-Lambda-G}) is redistributed 
to the vertex function in order to obtain a vertex function 
that is suitable for use with the elementary Dirac propagator.   
The remaining  $\sqrt{Z^{\pm}(p)/Z_2}$ factor in the propagators 
should be distributed to vertex functions preceding or following the
ones indicated in Eq.~(\ref{eq:G-Lambda-G}).

Figure~\ref{fig:F1++_off}
shows the variation with off-shell momentum p$^2$ for 
the F$_1^{++}(M^2, Q^2, p^2)$ form factor, with $p^2$ 
being the off-shell momentum.  A factor $\sqrt{Z^{(+)}(p)/Z_2}$
is included for the off-shell leg.
Similar results are obtained for the F$_2^{++}$ 
and  F$_5^{++}$ form factors.  Roughly, when p$^2$ varies from .8 M$^2$ 
to 1.2 M$^2$, the form factor varies from 0.8 to 1.4 
times the on-shell form factor.  
The off-shell variation of form factors is stronger than has been 
found in the work of Tiemeijer and Tjon\cite{TiemeijerTjon90} 
or that of Naus and Koch
\cite{NausKoch89}.

For couplings between $+$ and $-$ states, the form factors generally are off
shell because the momentum $p$ of the negative state differs 
from $W_p = -M$, where $W_p = \sqrt{p^2}$.  Typically, 
$+$ to $-$ couplings are evaluated near $W_p = + M$, and thus they 
should include a factor  $\sqrt{Z^{(-)}(p)/Z_2}$ from the 
negative-energy propagator in order to be compared with 
elementary couplings.  

Off-shell dependence of the F$_1^{+-}$, F$_2^{+-}$ 
and F$_5^{+-}$ form factors 
is different in each case.  It is shown in  
Figures~\ref{fig:F1+-_off}, {\ref{fig:F2+-_off} 
and {\ref{fig:F5+-_off}.  In each case, the F$^{+-}$
form factor is shown as the ratio to the on-shell F$^{++}$
form factor, and a factor $\sqrt{Z^{(-)}(p)/Z_2}$ is included.
The composite 
nucleon model gives nontrivial modifications of 
the form factors with off-shell momentum.    

Although the pure pseudoscalar operator $\gamma_5$ 
appears for each $\rho_f$ and 
$\rho_i$ value in the pion-nucleon vertex function, the form 
factors differ, i.e., F$_5^{+-} \neq ~$F$_5^{+-}$, as mentioned above.  
It is instructive to compare with an elementary vertex that contains a 
fraction $\lambda$ of pseudovector and $ 1 - \lambda$ of 
pseudoscalar couplings as follows, 
\begin{equation}
\Lambda_5 ^{\rm elem} (p_f,p_i) =  \lambda\gamma_5 
\frac{p\!\!\!/_i - p\!\!\!/ _f}{2M} + (1 - \lambda) \gamma_5 .
\end{equation}
Expanding by use of the projection operators and 
specializing to on-mass-shell kinematics
yields
\begin{eqnarray}
\Lambda^{\rm elem}_5 (p_f,p_i) =\sum_{\rho_f, \rho_i = \pm} 
L^{\rho_f} (p_f)\gamma_5 
\left[ \lambda  \frac{  \rho_f + \rho _i }{2}+ 1 - \lambda   
\right]  L^{\rho_i} (p_i).
\end{eqnarray}
On mass shell, the $++$ vertex is $\gamma_5$ independent of the mixing 
parameter $\lambda$.  The $+-$ vertex is proportional to 
$(1 -\lambda) $ and thus is suppressed 
for pseudovector coupling. 
A measure of the fraction of pseudovector coupling  
shows up in the ratio of $+-$ and $++$ form factors.  For the composite 
nucleon model, we define an equivalent pseudovector fraction in order
to give a simple interpretation of the 
different $++$ and $+-$ couplings as follows,  
\begin{equation}
\lambda =  1 - \left( \frac{F_5^{+-}(p',p)}{F_5^{++}(p',p)}
\sqrt{\frac{Z^{(-)}(p)}{Z_2} }
\right)_{W_{p'} = W_p = +M} .
\end{equation}  
Figure~\ref{fig:lambda} shows this ratio for the model nucleon.  In the low
Q range, the composite model produces 75\% pseudovector coupling of the pion
starting from a pseudoscalar coupling to the quark.  (If the 
factor $\sqrt{Z^{(-)}(p)/Z_2}$ were omitted, it would be 
94\% pseudovector.)  
At Q $\approx$ 1 GeV/c, the vertex becomes closer to pseudoscalar. 

To summarize this section, the model nucleon has realistic 
 charge and magnetic form factors.
The pion form factor is similar to the magnetic one and the $\pi$N vertex 
is about 75\% pseudovector and  25\% pseudoscalar.  Couplings 
between $+-$ and $++$ states differ, which
is a general feature of off-shell vertices.  

\section{Second-order interactions - the virtual photopion amplitude}

Using the couplings discussed in the previous section, 
we consider a virtual photopion 
production process.  This involves inserting 
a photon and a pion in all possible 
ways into the propagator and extracting the 
scattering amplitude as the residue
of the poles in G(p$_f$) and G(p$_i$) as before.  We also consider 
a standard hadronic treatment of the same process for comparison.

\subsection{Composite nucleon analysis}

For the process in which a nucleon with initial 
momentum p$_i$ 
absorbs a photon of momentum q, propagates with momentum p$_i$ + q,  
and subsequently emits a pion of
momentum r, ending up with momentum p$_f$, 
where  p$_i$ + q = p$_f$ + r,  the resulting amplitude is shown in
Fig.~\ref{fig:Vmu5} and is given by (omitting isospin factors)
\begin{eqnarray}
V_{5,\mu} (p_f,p_i&&+q,p_i) \equiv  
 \sum _{\rho} \bar{u} (p_f) g_{\pi} \gamma_5 F^{+,\rho} _{5}
(p_f,p_f+r) 
G^{\rho} (p_i+q) \widetilde{\Lambda} ^{\rho, +}
_{\mu} (p_i+q,p_i) u(p_i) .
\label{eq:V5mu}
\end{eqnarray}
For the crossed process in which the nucleon first   
emits a pion of momentum r and subsequently absorbs a photon of
momentum q, ending up with the same momentum p$_f$, the
amplitude is (omitting isospin factors)
\begin{eqnarray}
V_{\mu,5} (p_f,p_i+r,p_i) \equiv 
 \sum _{\rho} \bar{u} (p_f) \widetilde{\Lambda} ^{+,\rho} _{\mu}
(p_f,p_f-q) 
G^{\rho} (p_i-r) g_{\pi} \gamma_5 F^{\rho, +}
_{5} (p_i-r,p_i) u(p_i) 
\label{eq:Vmu5} 
\end{eqnarray}
These contributions to the photopion amplitude will 
be referred to as ``Born'' terms.

In the analysis, two factors of Z$_2$ arise, one 
from the external wave functions and 
another from the pole term of the propagator.  
These factors are absorbed into the two 
vertex functions so that all quantities appearing in 
Eq.~(\ref{eq:V5mu}) and 
(\ref{eq:Vmu5}) are renormalized. 
Renormalized photon vertex function $\widetilde{\Lambda}
^{\rho_f,\rho_i}_{\mu}$ is defined in 
Eq.~(\ref{eq:Lambdamu}) in terms of form factors F$^{\rho_f,\rho_i}_1$,
 F$^{\rho_f,\rho_i}_2$, and  F$^{\rho_f,\rho_i}_3$.
Propagation  
has been split into separate
factors for $\rho = +$ and $\rho = -$ states using covariant
projection operators.  Note that G$^+$ contains the nucleon pole term 
and the excited states, which in this case are quark 
and diquark scattering states. 
Similarly, G$^-$ is the negative-energy propagation that occurs in Z-graphs. 
However, the standard Z-graph is based on noncovariant 
projection of the propagator
and this causes some differences when nucleon momenta 
are not close to the mass shell.  
All of these elements arise also in a hadronic 
description.  

The variation of V$_{5,\mu}$ with momentum transfer 
is characterized roughly by F(q$^2$) G$^+$(p$_i$ +q) F(r$^2$), 
where F(q$^2$) is 
a typical form factor and G$^+$ is the positive-energy propagator.  
At large q$^2$ and r$^2$, these contributions become 
small owing to the form factors involved.  
A similar estimate holds for  V$_{\mu,5}$.  Excited states of the nucleon 
do little to alter this behavior because they 
involve form factors that typically fall faster with 
increasing momentum transfer than the nucleon's form factors.  

  In addition, there are contact-like terms as indicated in 
Fig.~\ref{fig:Vmu5} that differ from those that
arise in a hadronic description.  They correspond to the
two orders in which the photon and pion interact with a 
constituent fermion
within a single vertex. Omitting isospin factors, they are defined by,
\begin{eqnarray}
C_{5,\mu} (p_f,p_i+q,p_i) \equiv \bar{u} (p_f) \Biggr[ 
ig Z_2 \int \frac{d^4k}{(2 \pi)^4} 
S(p_f-k;m) 
g_{\pi}  \gamma_5 S(p_i+q-k;m)  
 \nonumber \\  \gamma _{\mu} S(p_i-k;m) D(k;\mu,\Lambda_1) \Biggr] u(p_i)
\nonumber \\ &&
\label{eq:C5muloop}
\end{eqnarray}
\begin{eqnarray}
 C_{\mu,5} (p_f,p_i-r,p_i) \equiv \bar{u} (p_f) \Biggr[
ig Z_2  \int \frac{d^4k}{(2 \pi)^4}
S(p_f-k;m) 
\gamma_{\mu} S(p_i-r-k;m)  
\nonumber \\   g_{\pi} \gamma _{5} S(p_i-k;m) D(k;\mu,\Lambda_1)
\Biggr] u(p_i). 
\label{eq:Cmu5loop}
\end{eqnarray}
Initial and final
states are on-shell positive-energy states, i.e., $p_i^2=p_f^2=M^2$ and
$\rho_i=\rho_f=+$.  In the generalized model with additional
Pauli-Villars regulators, $C_{5,\mu}$ and $C_{\mu, 5}$
terms become sums of terms of the form given in Eqs.~(\ref{eq:C5muloop})
and (\ref{eq:Cmu5loop}).
In each $\Sigma_b$ of Eq.~(\ref{eq:Sigmabsum}), the 
fermion propagator $S(p_f - k;m_n)$
is replaced by $S(p_f - k; m_n) g_{\pi} \gamma_5 
S(p_i + q - k; m_n) \gamma_{\mu} S(p_i - k; m_n)$ to 
obtain $C_{5,\mu}$, or by $S(p_f - k;m_n) \gamma_{\mu}
S(p_i - r; m_n) g_{\pi} \gamma_5 S(p_i - k; m_n)$
to obtain $C_{\mu,5}$.

   Finally, there is an amplitude that results from the 
photon coupling to the charged pion. This is referred to as the
pion-in-flight amplitude, and it takes the form 
\begin{eqnarray}
A^{\pi}_{\mu}(p_f,p_i) = g_{\pi} e \vec{\tau} \cdot T_3 
\hat{\phi} \bar{u}(p_f) \widetilde{\Lambda}_5 
(p_f,p_i) u(p_i) G_{\pi}(r-q) J^{\pi}_{\mu},  
\end{eqnarray}
where $e T_3$ is the charge operator for the pion, and 
\begin{equation}
J^{\pi}_{\mu} = 2 r_{\mu} - q_{\mu}.
\end{equation}

The total amplitude for photopion production is the sum of Born and
contact-like parts, with appropriate isospin factors included, and the
pion-in-flight term.
\begin{eqnarray}
A_{\mu} (p_f,q,p_i) = \vec{\tau} \cdot \hat{\phi} \hat{e} 
V_{5,\mu} (p_f,p_i+q,p_i) + 
 \hat{e} \vec{\tau} \cdot \hat{\phi} V_{\mu,5}(p_f,p_i-r,p_i) 
 \nonumber \\ 
+ \vec{\tau} \cdot \hat{\phi} \hat{e} C_{5,\mu} (p_f,p_i+q,p_i) + 
\hat{e} \vec{\tau} \cdot \hat{\phi} C_{\mu,5}(p_f,p_i-r,p_i) 
+ A^{\pi}_{\mu} (p_f,p_i) 
\label{eq:Apiphoton}
\end{eqnarray}
Note that the order of isospin factors is important as they do not commute.  

   Gauge invariance implies conservation of the EM current, viz.,
$q^{\mu} A_{\mu} = 0$ when the pion is on mass shell, i.e., 
$r^2 = m_{\pi}^2$.  When the pion is off mass shell, 
there is in general a nonzero result proportional to $G_{\pi}^{-1} (r) = 
r^2 - m_{\pi}^2$.  The required form is realized in the photo-pion
amplitude A$_{\mu}$ because of the following Ward-Takahashi
identities\cite{Ward50,Takahashi57}.
\begin{eqnarray}
 q ^{\mu}  V_{\mu,5} = 
 \bar{u} (p_f) g_{\pi} \widetilde{\Lambda}_5 (p_f, p_i + q) u(p_i)
\end{eqnarray}
\begin{eqnarray}
q ^{\mu} V_{5,\mu} = - \bar{u} (p_f) g_{\pi} 
\widetilde{\Lambda}_5 (p_i-r, p_i) u(p_i), 
\label{eq:V-WTI}
\end{eqnarray} 
\begin{eqnarray}
 q ^{\mu}  C_{\mu,5} = \bar{u} (p_f) g_{\pi} 
\widetilde{\Lambda}_5 (p_f, p_i) u(p_i)
- \bar{u} (p_f) g_{\pi} \widetilde{\Lambda}_5 (p_f, p_i + q) u(p_i)
\end{eqnarray}
\begin{eqnarray}
 q ^{\mu}  C_{5,\mu} =  \bar{u} (p_f) g_{\pi} 
\widetilde{\Lambda}_5 (p_i-r, p_i) u(p_i) 
 -  \bar{u} (p_f) g_{\pi} \widetilde{\Lambda}_5 (p_f, p_i) u(p_i).  
\label{eq:C-WTI}
\end{eqnarray} 
These identities may be derived by use of Eqs.~(\ref{eq:vertex-WTI}),
(\ref{eq:C5muloop}) and (\ref{eq:Cmu5loop}).  
In the contact-like terms, one needs to use the 
elementary Ward-Takahashi identity for Dirac propagators
\begin{equation}
q^{\mu} S(p+q;m) \gamma_{\mu} S(p;m) = S(p;m) - S(p+q;m) .
\end{equation}

  The pion-in-flight term is rewritten in terms of a 
commutator involving the nucleon's charge operator, $\hat{e}$, using the
isospin identity $e \vec{\tau} \cdot T_3 \hat{\phi} = - [ \hat{e},
\vec{\tau} \cdot \hat{\phi} ]$. 
Its contribution to the divergence of the amplitude then is
\begin{eqnarray}
 q^{\mu} A^{\pi}_{\mu} (p_f,p_i) =  \big[\hat{e}, 
\vec{\tau}\cdot\hat{\phi}\big] 
\bar{u} (p_f) g_{\pi} \widetilde{\Lambda}_5 (p_f, p_i) u(p_i)
 G_{\pi} (r-q) \big[ G^{-1}_{\pi} (r) - G^{-1}_{\pi} (r-q) \big]
\end{eqnarray}

Contributions to $q^{\mu} A_{\mu}$ from the Born terms are 
cancelled exactly by the contributions
from the contact-like terms that have the same isospin factors, and 
the remaining contributions from the contact terms are
cancelled by the second term from the pion-in-flight contribution.  This
leaves only a term proportional to 
$ G^{-1}_{\pi} (r)$ that vanishes for an on-shell pion.  The 
full amplitude is gauge invariant and the presence of the contact-like
terms is essential for this result.

The distinguishing feature of the contact-like terms is 
that no propagator for the composite system occurs between interactions.  
Thus, there is not a separate form factor for each interaction. 
However, the contact-like terms do depend 
upon the momentum transfer.  They differ from a 
form factor mainly by the presence of an extra fermion propagator 
in the loop integrals of Eqs.~(\ref{eq:C5muloop}) and (\ref{eq:Cmu5loop}). 
If the extra propagator lines were shrunk to a point, the contact-like
terms would be related to form factors at momentum transfer q-r.  This
suggests that the contact terms should behave like F((q-r)$^2$) s(q),
where s(q) accounts for the extra propagator.  Our calculations show that 
s(q) is given roughly by s(q) = $\kappa^2$/($\kappa^2$- q$^2$ ), where
$\kappa$ is a typical fermion mass.
Comparing with  V$_{5,\mu}$ and V$_{\mu,5}$,  the contact-like terms fall 
more slowly with increasing momentum transfer
and ultimately they dominate the scattering. 

\subsection{Elementary particle with form factors  analysis}

    A standard treatment of meson-exchange currents in nuclear
physics is to construct graphs corresponding to elementary
particles and then to insert form factors at 
the vertices\cite{Mathiot89,Riska89}.
The form factors are obtained from on-shell matrix elements, e.g.,
from phenomenological fits to 
electron scattering data for a free proton target.  
 
   Treating the composite nucleon in this way, 
there are Born contributions of the 
$V_{5,\mu}$ and $V_{\mu,5}$ types, which are
evaluated using the F$^{++}$ form factors, and the pion-in-flight term. 
We consider both pseudoscalar and pseudovector pion-nucleon coupling 
in the elementary
particle amplitude, and there is an additional 
contact term $ C_{\mu}^{Elem(PV)}$ 
in the pseudovector case that 
results from gauging the derivative of the pion field.

The elementary amplitude with pseudovector pion coupling is defined as,  
\begin{equation}
A^{\rm Elem}_{\mu} =  \vec{\tau} \cdot \hat{\phi} \hat{e} 
V_{5,\mu}^{\rm Elem} (p_f,p_i+q,p_i) +
 \hat{e} \vec{\tau} \cdot \hat{\phi} V_{\mu,5}^{\rm Elem}(p_f,p_i-r,p_i) 
+ C_{\mu}^{Elem(PV)} (p_f,p_i) + A^{\pi}_{\mu}(p_f,p_i),
\label{eq:A-elem}
\end{equation}
where
\begin{eqnarray}
V_{5,\mu}^{\rm Elem}  (p_f,p_i+q,p_i) \equiv 
  \bar{u} (p_f) g_{\pi} F ^{+,+} _{5} (r) 
\gamma _5 
 \left( \frac{p\!\!\!/_i +q\!\!\!/ + M }{2M} \right) 
 \frac{1}{ p\!\!\!/_i+q\!\!\!/ - M + i \eta}  
\nonumber \\ \times
\left[ F_1 ^{+,+}(q)   
\gamma _{\mu} + i \sigma _{\mu,\nu} q^{\nu}
F^{+,+} _2 (q) \right] u(p_i) .
\label{eq:V5muElem}
\end{eqnarray}
Similarly, the crossed contribution is 
\begin{eqnarray}
V_{\mu,5}^{\rm Elem} (p_f,p_i-r,p_i) \equiv  
  \bar{u} (p_f) \Bigr[  F_1 ^{+,+}(q)  
\gamma _{\mu} 
+ i \sigma _{\mu,\nu} q^{\nu} 
F^{+,+} _2 (q) \Bigr]    
\frac{1}{p\!\!\!/_i-r\!\!\!/ - M + i \eta}  
\nonumber \\ \times g_{\pi}  
\gamma_5 \left(  \frac{M +r\!\!\!/ - p\!\!\!/_i}{2M} 
\right) F ^{+,+} _{5}(r)
u(p_i),
\label{eq:Vmu5Elem} 
\end{eqnarray}
In Eq.~(\ref{eq:V5muElem}), a pseudovector vertex factor  
$(p\!\!\!/_i +q\!\!\!/ - p\!\!\!/_f)/2M$ has been evaluated by use of 
$ \bar{u} (p_f)	\gamma _5 p\!\!\!/_f = - \bar{u} (p_f) \gamma _5 M$. 
Similarly, in Eq.~(\ref{eq:Vmu5Elem}), a $p\!\!\!/_i $ 
in the pseudovector factor $(p\!\!\!/_i + r\!\!\!/ - p\!\!\!/_i )/2M$ has
been replaced by M by use of the Dirac equation.  When pseudoscalar pion
coupling is used, these factors are omitted.   

Note that the transition matrix elements to an intermediate
negative-energy state in Eqs.~(\ref{eq:V5muElem}) and(\ref{eq:Vmu5Elem}) 
are based upon the same form factor 
as for the on-shell transition to an intermediate positive-energy state
in the elementary amplitudes.  However, when the pion vertex is pseudovector
there is reduced coupling to the negative-energy states.  In the 
corresponding Born amplitudes of Eqs.~(\ref{eq:V5mu}) and (\ref{eq:Vmu5}),  
transitions are 
based upon off-shell vertex functions that differ in general for the two
transitions.  

   Hadronic contact terms are implied by the off-shell 
factors $(p\!\!\!/_i +q\!\!\!/ + M)/(2M)$ in
Eqs.~(\ref{eq:V5muElem}) and $(M + r\!\!\!/ - p\!\!\!/_i)/(2M) $ in 
(\ref{eq:Vmu5Elem}).  For example the first factor may be rewritten 
as $1 + (p\!\!\!/_i +q\!\!\!/ - M)/(2M)$,  where the numerator of the
second part cancels the propagator.  A corresponding
rearrangement applies to $V_{\mu,5}^{\rm elem}$.  As a 
consequence, the pseudovector vertex can be replaced by a pseudoscalar one
plus hadronic contact terms in which the factor $1/(2M)$ replaces the
nucleon propagator between the photon absorption and pion emission.
However, the form factors at the pion and electromagnetic 
vertices remain, which is why we refer to 
such terms as hadronic contact terms.
They are distinct from the contributions we refer 
to as contact-like terms in the composite model.

The resulting hadronic contact terms for pseudovector pion coupling
have parts in which the $F_1^{++}$ form factor appears at the 
electromagnetic vertex.  These contact terms exactly cancel 
with the $C^{Elem(PV)}_{\mu}$ term.  This
leaves only the parts of hadronic contact terms that involve 
the magnetic form factor $F_2^{++}$.  For pseudoscalar pion coupling, the
situation is simpler because neither the contact terms from the off-shell
vertices nor the one from gauging the derivative of the pion field are
present. Consequently, there is a near equivalence of the 
pseudoscalar and pseudovector elementary amplitudes.  
After the cancellations in the pseudovector elementary amplitude, the 
only surviving differences from the pseudoscalar elementary amplitude are 
the parts of hadronic contact terms involving $F_2^{++}$.  

Vertex functions in the elementary amplitude cannot be defined 
precisely because of a basic conflict between the
use of on-shell form factors and the conservation of four-momentum.  
Except when q happens to be equal to the difference of 
two on-shell momenta, one
cannot have an on-shell vertex.  In the elementary Born amplitudes of
Eqs.~(\ref{eq:V5muElem}) and (\ref{eq:Vmu5Elem}) 
we have used on-shell form factors at vertices.  This is consistent with   
having a vertex function that does not depend 
on $p_i^2$ or $p_f^2$ and therefore can be evaluated with on shell
initial and final momenta whose difference is the momentum transfer,
i.e., $p_i = (E_Q,0,0,-Q/2)$ and  $p_i = (E_Q,0,0,Q/2)$, where E$_Q$ =
$\sqrt{ {\rm M}^2 + {\rm Q}^2/4}$, 
even though these are not the four-momenta that occur in the 
process.  The assumption that the vertex depends only on $Q^2$,
not the off-shell momenta, 
is often used when the off-shell 
dependence of the vertex function is unknown,  
but it does not have a sound theoretical basis for a composite particle.
 
A standard nonrelativistic analysis would be similar to the 
elementary analysis described above.  Relativistic kinematics would 
be used but the Z-graph parts of amplitudes would be omitted 
and typically a pseudoscalar pion coupling would be used.  
Calculations based on such a definition of a nonrelativistic 
amplitude will be discussed in Sec. 6.

To summarize this section, the composite nucleon model has features 
which are similar to those of a hadronic theory, which also has, 
at least in principle, vertex functions that are off shell, and that are
different functions for the different $\rho$-spins.  Because the
off-shell vertex functions are not known, the standard hadronic analysis
uses the on-shell form factors in their place.   
The main feature that distinguishes the composite 
particle analysis from an elementary 
particle analysis is the presence of contact-like terms.  They 
describe scattering from the 
partons and are related to 
off-forward parton distributions.     

\section{Deep inelastic scattering}

In order to obtain a rough normalization of the contact-like 
terms, we consider 
deep inelastic scattering from the composite nucleon.  
It is characterized by a hadronic tensor 
that takes a gauge-invariant form as follows \cite{West75,Roberts90},
\begin{eqnarray}
\frac{W^{\mu \nu} }{4 \pi M} = -\left( g^{\mu \nu} 
- \frac{q^{\mu} q^{\nu}}{q^2} 
\right) W_1 + \left( p^{\mu} - q^{\mu} \frac{p \cdot q}{q^2} \right)
 \left( p^{\nu} - q^{\nu} \frac{p \cdot q}{q^2} \right) \frac{W_2}{M^2} ,
\end{eqnarray}
Structure functions W$_1$ and W$_2$ depend on two scalar invariants:  
Q$^2$ = $-$q$^2$, and $\nu$ = p $\cdot$ q/M.  
In the limit Q$^2 \rightarrow \infty$, 
with x $\equiv$ Q$^2$/(2M $\nu$) held fixed, these 
functions become dependent only on x as follows \cite{CallanGross69},
\begin{equation}
M W_1(x,Q^2) \rightarrow F_1(x) = {1 \over 2} f(x),
\end{equation}
and 
\begin{equation}
\nu W_2(x,Q^2) \rightarrow F_2(x) = x f(x).
\end{equation}
This scaling behavior is a 
consequence of scattering from 
point-like constituents of the nucleon, 
with f(x) being the the probability 
of scattering from a parton that carries a fraction 
x of the nucleon's momentum.  

A simple way to obtain the forward parton distribution f(x) is to 
calculate W$^{xx}$. Either in the lab frame, where p = (M, 0, 0, 0,) and 
q = ($\nu$, 0, 0, $\sqrt{Q^2 + \nu^2}$), 
or in the c.m. frame of the final state, 
the x-components of q and p vanish.  For finite Q we have
\begin{equation}
2 M W_1(x,Q^2) = \frac{W^{xx}}{2 \pi},
\label{eq:2MW1}
\end{equation}
and in the asymptotic limit
\begin{equation}
 f(x) =   \lim_{Q^2 \to \infty} 2 M W_1(x,Q^2) .
\label{eq:fofx}
\end{equation}

Batiz and Gross \cite{BatizGross99} have analyzed the scaling limit for
a composite nucleon model that is essentially similar to the one used in
this paper.  Their analysis is for 
one space and one time dimension.  They show that scaling in a general 
gauge involves a cancellation between a gauge-dependent part of the impulse 
approximation graph of Fig.~\ref{fig:distr} and a gauge-dependent part of
the final-state interaction.  This is related to the 
Ward identities of Eqs.~(\ref{eq:V-WTI})
and (\ref{eq:C-WTI}) which imply that gauge invariance requires
cutting both the Born and contact-like terms.  
Using the  Landau prescription, Batiz and Gross
split the impulse amplitude 
into a gauge-invariant part and a remainder.
The gauge-invariant 
part of the impulse graph provides the scaling result.
The gauge variant remainder cancels with part of the 
final-state interaction such that the resultant 
contribution of these parts vanishes 
at least as fast as 1/Q$^2$.  
In this section, we follow Batiz and Gross by using the 
Landau prescription for 
three space dimensions and one time dimension.  The results differ
because of integrals over the angles of final state particles and 
because the phase space in 3D differs 
from that in 1D.     

The hadronic tensor based on the impulse graph of Fig.~\ref{fig:distr} 
is calculated in the c.m. frame
of the final state,
\begin{equation}
W^{\mu \nu} = \frac{1}{2} \sum_{s,s_1} \frac{|\vec{p}_1|}{4 \pi W}
\frac{1}{4 \pi} \int d \Omega _1 {\cal{T}}^{\mu \dag}  {\cal{T}}^{\nu}  ,
\end{equation}
where W is the total energy of the final state that contains an on-shell 
quark of momentum p$_1$ = (E$_1$, {\bf p}$_1$) and 
an on-shell boson of momentum
(W - E$_1$, -{\bf p}$_1$). Amplitude ${\cal{T}}^{\mu}$ describes 
the impulse approximation graph for 
scattering from the fermion constituent.  Using the Landau prescription as 
in Ref.~\cite{BatizGross99} to obtain a gauge-invariant current, this is 
\begin{eqnarray}
{\cal T}^{\mu} = \frac{\sqrt{| g Z_2| 4mM}}{(p_1 -q)^2 - m^2} 
\bar{u}_1(p_1;m) \Bigr[ 2 p_1^{\mu} - \frac{2p_1 \cdot q}{q^2} q^{\mu} 
- \gamma^{\mu}
 q \!\!\!/ ~+q^{\mu} \Bigr] u(p;M),
\end{eqnarray}
where u$_1$(p$_1$;m) is a Dirac spinor for the quark of mass m and 
u(p;M) is a Dirac spinor for the nucleon of mass M.  
The factor $\sqrt{|g Z_2|}$u(p;M)  
is the vertex function for the nucleon to fragment into a quark and a boson.  

An equivalent form of the hadronic tensor, which we use, is
\begin{equation}
W^{\mu \nu} = \frac{|\vec{p}_1|}{4 \pi W}
\frac{|g Z_2|}{4 \pi} \int d \Omega _1 \frac{1}{[(p_1 - q)^2 - m^2]^2}
{\cal M}^{\mu \nu}, 
\end{equation}
where we define
\begin{eqnarray}
{\cal M}^{\mu \nu} = \frac{1}{2} Tr \left\{ \left( 
2 p^{\mu}_1  - \frac{2p_1 \cdot q}{q^2} q^{\mu} 
- \gamma^{\mu} q \!\!\!/ ~+q^{\mu}
\right) (p\!\!\!/~+M) \right. 
\nonumber   \\ \times
 \left. \left( 
2 p^{\nu}_1  - \frac{2p_1 \cdot q}{q^2} q^{\nu} 
- \gamma^{\nu} q \!\!\!/ ~+q^{\nu}
\right) (p\!\!\!/_1+m) \right\} . 
\end{eqnarray}
Specializing to ${\cal M}^{xx}$, we find
\begin{eqnarray}
{\cal M}^{xx} = 2( 4p_1^x p_1^x + Q^2) (p \cdot p_1 + M m ) 
+ 4 M \nu (q \cdot p_1 - 2 p_1^xp_1^x ),
\end{eqnarray}
where terms not involving x-components have arisen 
from use of $\gamma^{x}\gamma^{x} = -1$. 
Expressing the vectors in the c.m. frame, where 
p$_1$ = (E$_1$, $|{\bf p}_1|$sin$\theta_1$cos$\phi_1$, 
 $|{\bf p}_1|$sin$\theta_1$sin$\phi_1$ , $|{\bf p}_1|$cos$\theta_1$ ), 
leads to the following expression
\begin{eqnarray}
{\cal M}^{xx} =  
 8 |{\bf p}_1|^2 sin^2\theta_1cos^2\phi_1
\Big[p^0 E_1 - p^z |{\bf p}_1| cos\theta_1
+ M m 
- M \nu\Big] 
 \nonumber \\ 
+  4 M \nu\Big[ q^0 E_1 - q^z  |{\bf p}_1| cos\theta_1 \Big]
+2Q^2\Big[p^0 E_1 - p^z |{\bf p}_1| cos\theta_1 +M m\Big] 
\end{eqnarray}
Carrying out the elementary 
angle integrations produces,
\begin{equation}
\frac{W^{xx}}{2 \pi} =\frac{| g Z_2|| {\bf p}_1|}{8 \pi^2 W} 
\left( \frac{C_0}{a^2 - b^2} + 
C_1 \ln\left( \frac{a + b}{a - b} \right) + C_2 \right) ,
\label{eq:Wxx} 
\end{equation}
\begin{eqnarray}
a &=& -Q^2 - 2 q^0E_1 
 \nonumber \\ &&
\rightarrow \frac{-Q^2}{2 x} + \frac{2 x - 1}{2(1 - x)}[m^2 - \mu^2] 
\end{eqnarray}
\begin{eqnarray}
b &=& 2 q^z  |{\bf p}_1|
 \nonumber \\ &&
\rightarrow \frac{Q^2}{2 x} + \frac{2M^2 x(1-x) - m^2 - \mu^2}{2(1-x) }
\end{eqnarray}
\begin{eqnarray}
C_0&&= 4 M \nu q^0 E_1 + 2Q^2(p^0E_1 + M m) \nonumber \\ &&+ [4 M \nu q^z  
|{\bf p}_1| + 2Q^2 p^z  |{\bf p}_1|] \frac{a}{b} 
 \nonumber \\ &&
\rightarrow \frac{Q^2}{x} [ ( M x + m )^2 - F_{\mu}(x) ]  
\end{eqnarray}
\begin{eqnarray}
C_1&& = 4  |{\bf p}_1|^2 (p^0E_1 + M m - M \nu) \frac{a}{b^3} 
- 2  |{\bf p}_1|^3 p^z \frac{1}{b^2}
\left(1 - \frac{3 a^2}{b^2}\right) 
- (2 M \nu q^z  |{\bf p}_1| + 
Q^2 p^z |{\bf p}_1|) \frac{1}{b^2} 
 \nonumber \\ &&
\rightarrow x-1
\end{eqnarray} 
\begin{eqnarray}
C_2 =&& -8 |{\bf p}_1|^2 (p_0E_1 + M m - M \nu) \frac{1}{b^2} - 
12 |{\bf p}_1|^3 p^z \frac{a}{b^3}
 \nonumber \\ &&
\rightarrow x-1 
\end{eqnarray}
and we have defined 
\begin{equation}
F_{\mu}(x) = m^2 (1-x) + \mu^2 x -M^2 x (1-x).
\end{equation}
Furthermore, the phase-space factor approaches a constant, 
\begin{eqnarray}
\frac{ |{\bf p}_1|}{ W} \rightarrow \frac{1}{2}.
\end{eqnarray}
In the expressions given above, the limiting form as  
 Q$^2 \rightarrow \infty$ with x = Q$^2$/(2 M $\nu$) 
fixed is indicated following the arrow.
We have used a number of kinematical relations that can be found in the 
paper of Batiz and Gross. 

Note that ln((a$+$b)/(a$-$b)) arises in the structure function.  
Because a$+$b $\rightarrow -$
F$_{\mu}$(x)/(1 $-$ x)
is independent of Q, but 
a $-$ b $\rightarrow -$Q$^2$/x is not, there is 
a $\ln$(Q$^2$) term in W$^{xx}$.
This is cancelled when the Pauli-Villars subtraction is 
made, i.e., when the parton 
distribution is calculated as 
the discontinuity of the subtracted bubble graph.  
Clearly, scaling in 3+1 dimensions depends
upon the subtraction, which was not the case in the 1+1 
dimensional analysis of Ref.~\cite{BatizGross99}.
>From Eqs.~(\ref{eq:Wxx}),(\ref{eq:2MW1}) and (\ref{eq:fofx}), 
we find the following parton distribution for 
the bubble graph with a fermion of mass m, a boson of 
mass $\mu$, and a Pauli-Villars subtraction of mass $\Lambda_1$.
\begin{eqnarray}
f_b(x;m,\mu,\Lambda_1) = \frac{| g Z_2| (1-x)}{16 \pi^2} 
\Biggr\{  (Mx+m)^2
\left( \frac{1}{F_{\mu}(x)} -  \frac{1}{F_{\Lambda_1}(x)} \right)  
-\ln \Biggr( \frac{F_{\mu}(x)}{F_{\Lambda_1}(x)}   
\Biggr) \Biggr\}
\end{eqnarray} 

For the case in which additional subtractions are made, the parton 
distribution becomes a linear combination of terms as in 
Eq.(~\ref{eq:Sigmabsum}), i.e.,  
\begin{eqnarray}
&&f(x) =~~~ f_b (x;m,\mu, \Lambda_1) 
+~ \beta f_b(x;m,\Lambda_1,\Lambda_2) 
 \nonumber \\ &&
-\alpha \Big[ f_b (x;m_1,\mu, \Lambda_1) 
+ \beta f_b(x;m_1,\Lambda_1,\Lambda_2) \Big] 
 \nonumber \\ &&
-(1 - \alpha)\Big[ f_b (x;m_2,\mu, \Lambda_1) 
+ \beta f_b(x;m_2,\Lambda_1,\Lambda_2) \Big]
\label{eq:fbsum}
\end{eqnarray}

Figure~\ref{fig:Wxxplot} shows the inelastic 
structure function 2MxW$_1$(x,Q$^2$)
and its limit as Q$\to \infty$, xf(x),  
for the same parameters 
that allow a reasonable description of the nucleon 
form factors.  For finite Q, the
energy transfer $\nu = Q^2/(2Mx)$ also is finite.  
It must be greater than the total mass 
of the constituent quark and diquark
for each combination that enters Eq.~(\ref{eq:fbsum}) 
in order to avoid spurious threshold effects.
This restricts x not to be too close to 1.  For 
finite Q, we show W$_1(x,Q)$ only for $\nu$ 
greater than about 200 MeV above threshold. 
The solid line in Fig.~\ref{fig:Wxxplot} shows the asymptotic 
limit Q$\to \infty$, which is xf(x).  
Already by Q$\approx$ 2 GeV/c, the inelastic structure function is 
close to its asymptotic limit for our quark-diquark model of a nucleon.
Our result for $xf(x)$ is more peaked than that obtained recently by 
Mineo, Bentz and Yazaki \cite{MineoBentzYazaki99} from 
consideration of a three-quark 
model, and both results lack sufficient strength near x = 0
in comparison with experimentally determined parton distributions. 
Reference \cite{MineoBentzYazaki99} considers what one should 
expect for xf(x) at low Q.
Using QCD evolution to relate high and low Q, the 
nucleon's parton distribution 
is found to be more peaked at low Q, and not 
unlike the xf(x) that we find, with a
peak near x = m/M, where m is the lightest 
fermion mass.  However, our results at Q$^2$
= 4 (GeV/c)$^2$ are more peaked than the results 
of Ref. \cite{MineoBentzYazaki99} 
at Q$^2$ = 0.16 (GeV/c)$^2$.  
For x $>$ .6, xf(x) in Figure~\ref{fig:Wxxplot} 
is negative owing to the fact that the fermionic subtractions 
are not hermitian.  This 
is a deficiency of the model used.  Our results 
for the parton distribution are influenced by the choice of subtractions 
that have been incorporated in order to obtain a good description of the 
nucleon's form factors.  We have given preference to obtaining
a realistic form factor in selecting parameters. 

Owing to the normalization factor $|gZ_2|$, the parton distribution 
automatically is normalized according to 
\begin{equation}
\int _0 ^1 dx f(x) = 1.
\label{eq:fofxnorm}
\end{equation}
See Ref.  \cite{BatizGross99} and Appendix A in this regard.  
However, only the fermion constituent has charge and the momentum sum rule
is, 
\begin{equation}
\int _0 ^1 dx\ x f(x) = .304. 
\end{equation} 
Thus, 70\% of the momentum is carried by the diquark in this model.

To summarize this section, the model for a nucleon as a 
bound state of a quark and diquark is consistent with scaling 
in deep inelastic scattering.  The normalized parton distribution  
provides a rough normalization for the 
contact-like terms in the large Q limit.    

\section{Calculations for meson-exchange current amplitude}

   Virtual photopion production amplitude $A^{\mu}$ that has been 
defined in Eq.~(\ref{eq:Apiphoton}) and discussed 
in Sec. 4 contributes to the electromagnetic current in electron-deuteron 
scattering. For this case, the emitted pion is absorbed on a second nucleon
and, in general, there is a loop integration involving the pion momentum
and the deuteron wave functions of initial and final states.
For electron-deuteron scattering, the loop integration 
receives important contributions from the quasifree kinematics indicated
in Figure \ref{fig:QFkinematics}.  Each nucleon in 
the deuteron, only one of which is shown, has initial momentum
 $p_i = {1 \over 2}$P $-{1 \over 4}$q.  Figure \ref{fig:QFkinematics} 
shows the nucleon which absorbs a photon of momentum q and emits a pion of 
momentum $r = {1 \over 2}$q, ending up with momentum $p_f = {1 \over 2}$P
$+{1 \over 4}$q.  The second nucleon, not shown, also has 
initial momentum  $p_i = {1 \over 2}$P $-{1 \over 4}$q.
When the pion is absorbed on the second nucleon, its final momentum also
becomes $p_f$.  This process begins and ends with the two nucleons at zero
relative momentum.  It is favored because the 
deuteron wave function is largest at zero relative momentum.  
Pion-in-flight terms vanish for the selected kinematics, since $2 r_{\mu}
- q_{\mu} = 0$. They are omitted from our calculations. 
A calculation using deuteron wave functions is planned for a future work. 
For now, we focus on the quasi-free photopion amplitude and simply 
vary the momentum of the space-like virtual photon: q = (0, 0, 0, Q).

Although there are sixteen helicity amplitudes
$A^{\mu}_{\lambda_f,\lambda_i}$, 
for the quasifree kinematics with collinear momenta that we consider
only three amplitudes are significant.  An isospin-nonflip amplitude, a,
occurs in the time-component of the photopion amplitude
as follows,
\begin{eqnarray}
A^0_{\lambda_f,\lambda_i} = a \big\{ \hat{e}, 
\vec{\tau} \cdot \hat{\phi} \big\}
\chi_{\lambda_f}^{\dagger}   \sigma_z \chi_{\lambda_i},
\end{eqnarray}
where the isospin factor involves an anticommutator. 
Two isospin flip amplitudes, b and c, occur in the space-vector 
parts of the photopion amplitude as follows,
\begin{eqnarray}
A^{\pm}_{\lambda_f,\lambda_i} =
 b \big[ \hat{e}, \vec{\tau} \cdot \hat{\phi} \big] 
\chi_{\lambda_f}^{\dagger}  \sigma_{\pm} \chi_{\lambda_i} ,
\end{eqnarray}
\begin{eqnarray}
A^3_{\lambda_f,\lambda_i} = 
 c \big[ \hat{e}, \vec{\tau} \cdot \hat{\phi} \big]
\chi_{\lambda_f}^{\dagger}  \sigma_z \chi_{\lambda_i} ,
\end{eqnarray}
where the isospin factors involve a commutator.   

Although we calculate only the photopion amplitude, its 
role as a meson-exchange current in electron-deuteron 
scattering is of interest.  In that case, the isospin wave functions
$\hat{\phi}$ for the pion are replaced by 
the isospin operators $\vec{\tau}_2$ for the second nucleon.
The isospin-nonflip amplitude, a, is the only one that contributes as 
a meson-exchange current in elastic electron-deuteron scattering.  
However, for breakup of the deuteron, both isospin nonflip and 
isospin-flip amplitudes contribute.

\subsection{Isospin nonflip amplitude: a}

Figure \ref{fig:VGVandCEst} shows the absolute value of 
Born and contact-like contributions to amplitude $a$
in comparison with simple estimates of these contributions
suggested in Sec. 4:  
${\rm V}_{0,5} + {\rm V}_{5,0} 
\approx  {\rm C}_1{\rm Q F}_{\rm dipole}({\rm q}^2) {\rm G}^+
({\rm (p}_i +{\rm q} )  {\rm F}_{\rm dipole}
({\rm(r}^2)$  and 
${\rm C}_{0,5} + {\rm C}_{5,0}\approx$ C$_2$Q F((q-r)$^2$) S(q),  where
C$_1$ and C$_2$ are constants and s(q) = $\kappa^2$/(Q$^2$ + $\kappa^2)$.  
We find that $\kappa^2$ = .20 (GeV/c)$^2$ 
describes the Q dependence caused by the extra quark propagator in the  
contact-like terms.   A factor Q is included in the estimates because
there is such a factor in the amplitude for kinematical reasons.  
The point of this comparison 
is to show that the contact-like contributions are decreasing with Q 
faster than a form factor, as expected.  They nevertheless can be dominant 
when Q $>$ 1 GeV/c because the Born terms decrease 
even faster.  There is a zero in the Born amplitude that is not
reproduced in the estimate.  However, the estimate is quite good 
for individual covariant amplitues that go into the helicity matrix element 
(see Appendix C for the definition of covariant amplitudes). 

In Fig. \ref{fig:aBornFullElem_PS_NR}, we show $|a|$ for the 
Born (long dash line) and full amplitudes (solid line) 
for the composite nucleon.  Also shown (dash line) 
is the elementary amplitude $|a|$
that is based upon pseudoscalar pion coupling.  
Finally, we show (dotted line) a nonrelativistic 
amplitude that is based on pseudoscalar pion coupling and
the standard positive-energy propagator:
$\Lambda^+({\bf p}) /(p^0 - \sqrt{M^2 + {\bf p}^2})$.  Thus, 
the nonrelativistic amplitude differs from the elementary 
one by omission of the Z-graph part.  
Form factors used in the elementary particle and nonrelativistic
analyses are based on the quark-diquark model except that on-shell
++ form factors are used.
For small Q,
the Born contributions dominate for all cases because there is 
a pole in the intermediate nucleon propagator.  In the
vicinity of Q $=$ 1.2 GeV/c, two amplitudes that involve an intermediate 
propagator for a nucleon, i.e., the
the Born and nonrelativistic amplitudes, 
pass through zero.  They are negative at higher Q and their
magnitude is 40\% to 10\% of the full amplitude at Q $=$ 3 GeV/c.
The elementary amplitude based on pseudoscalar pion coupling has a 
zero near Q $\approx$ 2.8 GeV/c.  It provides the best approximation 
to the results of the composite model but is much smaller 
in magnitude for Q $\ge$ 1.5 GeV/c.  The amplitude for the 
composite nucleon model is dominated at large Q by the contact-like
terms, which do not change sign.  
Although the nonrelativistic amplitude 
tends at large Q to a magnitude similar to that of the full 
amplitude, it has the opposite sign and is not a 
useful approximation to the full result.  

Figure~\ref{fig:aElem_PS_ZandCont} shows the contact-like amplitude of the 
composite model, (solid line). The part of the Born amplitude that
comes from excited states and Z-graphs is shown by the long dash line.
It has been calculated by evaluating 
Eqs.~(\ref{eq:Vmu5}) and (\ref{eq:V5mu}) with the positive-energy propagator, 
$\Lambda^+({\bf p}) /(p^0 - \sqrt{M^2 + {\bf p}^2})$ and then subtracting
that result from the Born amplitude based on the full propagator of the 
composite model.  Next we show by the dash dot line  
the sum of parts of the composite nucleon 
amplitude that do not come from the Born terms
with the positive propagator, i.e., the sum of contact-like parts,
excited-state parts and Z-graph parts.  
Thus, the dot-dash line shows the sum of the 
amplitudes used in the solid and 
long-dash lines.  
The dash line shows the Z-graph part of the elementary amplitude 
based on pseudoscalar pion coupling.  
The excited-states-plus-Z-graph part of the composite-nucleon 
Born amplitude (long dash line in Fig.~\ref{fig:aElem_PS_ZandCont}) is
larger than the Z-graph part of the elementary Born amplitude (dashed
line) at low Q, but it decreases rapidly with Q.   
Because a less point-like structure for the composite nucleon would be
expected to provide even smaller contributions at large Q 
from excited states and Z-graphs, the smallness of the 
Born contributions of the composite model at large Q is 
expected to hold more generally.  The Q-dependence of the
pseudoscalar Z-graph contribution (dashed line) is notable for its
similarity to that of the contact-like contribution of the composite model
(solid line).

Because the pion vertex of the composite model 
is about 75\% pseudovector,   
we consider next the same set of comparisons using an elementary 
amplitude in which the pion coupling is pseudovector. 
In this case, we also include the contact term that is implied by 
gauging the derivative of the pion field, $C^{Elem(PV)}_{\mu}$.  
Figure~\ref{fig:aBornFullElem_PV_NR}  
shows that the elementary amplitude based on
pseudovector pion coupling (dashed line) provides a poorer 
approximation to the composite nucleon result (solid line).
This is because it has a zero near 1 GeV/c and has the wrong sign at 
large Q.  Figure~\ref{fig:aElem_PV_ZandCont} shows the difference between
the pseudovector elementary amplitude and the nonrelativistic amplitude
that uses pseudoscalar pion coupling by the dashed line.  

    We find that the use of pseudoscalar pion coupling in 
the elementary amplitude provides a better  
approximation to the amplitude of the composite model.  This 
is because it has a large Z-graph contribution that approximates the 
contact-like contribution of the composite model. As mentioned,
the pseudoscalar and pseudovector pion couplings produce different 
results only because of the magnetic couplings of the photon.  If 
the photon were to couple only via the charge current, 
$\gamma^{\mu} F_1(Q)$, the pseudoscalar and pseudovector 
elementary amplitudes that we consider would be equal.  However, the
magnetic part of the charge current, $\sigma^{\mu \nu}q_{\nu} F_2(Q)$,
changes this.   For pseudoscalar coupling, the Z-graph 
amplitude shown by the dashed line in Figure~\ref{fig:aElem_PS_ZandCont}
is proportional to $F_1(Q) + F_2(Q) = G_M(Q) $, 
whereas for pseudovector coupling, the dashed line in 
Figure~\ref{fig:aElem_PV_ZandCont} is  
proportional to $F_1  - F_2(Q)Q^2/(4 M^2) = G_E(Q)$.
The effect 
of the magnetic parts explains the different results in these graphs. 

Low-energy 
theorems that apply to the photopion amplitude at low Q
arise from chiral invariance.  Because the composite nucleon model
has essentially pseudovector pion coupling, which is consistent with
chiral invariance, one might expect the 
pseudovector elementary amplitude to provide a better approximation 
to the composite nucleon results.  This 
expectation fails at large Q because of the important contributions of
contact-like terms.   

\subsection{Isospin-flip amplitudes: b and c}
 
   Isospin-flip amplitudes $|b|$ and $|c|$ are shown in 
Figures~(\ref{fig:bBornFullElem_PS_NR}) and (\ref{fig:cBornFullElem_PS_NR}).
These amplitudes do not exhibit a pole at $Q = 0$ like the one in the 
isospin-nonflip amplitude, a.  Each vertex in the b and c amplitudes has 
a factor Q, thus cancelling the 
$1/Q^2$ from the propagator.  Consequently, the isospin flip amplitudes
are much smaller than a at small Q, but they can be comparable at large Q.
Elementary amplitudes based upon pseudoscalar pion coupling and pseudovector pion 
coupling are equivalent for the b and c amplitudes, and thus we show only
the pseudoscalar elementary amplitude.   This equivalence results because 
the hadronic contact terms involving the $F_2$ electromagnetic form factor
give a vanishing contribution for the kinematics that we consider.    

Figure~(\ref{fig:bBornFullElem_PS_NR}) shows that Born and nonrelativistic 
results for $|b|$ are very close to one another at small Q.  However, the 
Born amplitude is significantly smaller at larger Q.  The full composite
model result is close to that of the elementary amplitude
over the entire range of Q.  Born and nonrelativistic results both 
omit Z-graphs, whereas the full and elementary 
results both include Z-graphs.  It is apparent that the Z-graphs make a
significant contribution at Q $=$ 0, lowering the full and elementary 
results in comparsion with the Born and nonrelativistic ones.  
 
Contact-like terms of the composite nucleon cause the difference between
full and long-dash lines: these are significant but not  
dominant in the way they are for the isospin-nonflip amplitude, a. 

Figure~(\ref{fig:cBornFullElem_PS_NR}) shows that the full composite
model provides a much larger result for $|c|$ than is obtained from the
Born, elementary or nonrelativistic amplitudes.  Thus, $|c|$ and $|a|$
amplitudes show dominance of contact-like contributions at large Q, but
$|b|$ does not.  

It is clear that one would like to have better control of 
the normalization of contact-like terms in order to determine
the transition point where they may become dominant
contributions to electron scattering from nuclei.  However, 
the present model suggests that this could be near 1 GeV/c for 
the isoscalar MEC appropriate to elastic electron-deuteron scattering,
based upon the strong dominance of contact-like contributions to $|a|$.  
For the isospin-flip MEC contributions, which are relevant to 
electrodisintegration of the deuteron, each of the amplitudes a, b and c
contributes. A more complete calculation is required to see if the
contact-like terms may dominate the MEC at large Q.  
 
Use of pseudoscalar pion coupling improves the agreement with 
results of the composite model significantly for $|a|$, and is 
equivalent to pseudovector pion coupling for $|b|$ and $|c|$.  
The fact that pseudoscalar pion coupling seems to work fairly well
is not because it provides a description of the underlying physics,
which requires consideration of scattering from the quarks at large Q.  

\section{Conclusion}

A simple model of a composite nucleon is developed in which a
fermion and a boson, representing quark and diquark constituents
of the nucleon,
form a bound state owing to a contact interaction. 
Photon and pion couplings to the quark provide vertex functions
for the photon and pion interactions with the composite nucleon. 
By introducing and exploiting cutoff parameters of the Pauli-Villars
type, realistic electromagnetic form factors are obtained for the proton.
When a pseudoscalar pion-quark
coupling is used, the pion-nucleon coupling is 75\% pseudovector.
The small quark mass produces a vertex
behavior close to that expected from chiral invariance.

A virtual photopion amplitude is considered in which there are two
types of contributions: hadronic
contributions where the photon and pion interactions have an intervening
propagator of the nucleon, or its excited states, and 
contact-like contributions where the photon and pion 
interactions occur within a single vertex.  
Relative normalization of the two types of contribution 
is controlled by Ward-Takahashi 
identities at low momentum transfer.  
At high momentum transfer, scaling behavior is obtained for the 
composite nucleon already by Q $\approx$ 2 GeV/c.  
This provides a rough normalization of the contact-like parts because
the parton distribution is normalized (see Eq.~(\ref{eq:fofxnorm})).  
However, our model of a composite nucleon as a 
bound state of a quark and diquark 
yields a parton distribution that is 
peaked near x $=$ m/M, the ratio of quark to nucleon mass,
whereas the data suggest much less peaking and more strength at low x
values than the model gives.  

Calculations for the virtual photopion amplitude 
are performed using kinematics appropriate to its occurrence as 
a meson-exchange current in electron-deuteron scattering.
The results show that the contact-like terms dominate the meson-exchange
current for Q $>$ 1 GeV/c for the case of elastic electron-deuteron
scattering.  As Q increases, the dominance of the contact-like terms over
the Born terms of the composite nucleon can become very large,
suggesting that hadronic processes become unimportant when this occurs.
Our results indicate that contact-like terms
still have substantial Q dependence when they become dominant.

For the inelastic electron deuteron scattering, both 
isospin-nonflip and isospin-flip 
parts of the photopion amplitude can contribute.  Two of the three 
contributing amplitudes are dominated at large Q by contact-like terms
and the other is not.  A more complete calculation using deuteron wave
functions is needed in order to understand the role of 
contact-like contributions in deuteron breakup.

Off-shell effects in the hadronic vertex functions 
are found to be significant in the composite model.
They cause a significant suppression of Born 
contributions to the virtual photopion amplitude for Q $\ge$ 1 GeV/c.
This result is model-dependent, but it suggests that use of on-shell form
factors could be a poor approximation for momenta 
that are significantly off the mass shell.

An elementary amplitude based upon pseudovector pion coupling 
fails to provide a useful approximation to the full result 
of the composite model for the isospin-nonflip amplitude. 
This can be improved somewhat by using pseudoscalar pion 
coupling in the elementary amplitude.  The increased Z-graph contribution
gives a better approximation to the contact-like terms of the composite 
nucleon, but not to the underlying physics.     

Compositeness requires contact-like terms in
second-order interactions.  They have a direct connection to
off-forward parton distributions and can dominate
the scattering at large Q as they contain the leading 
partonic scattering process.  
Hadronic form factors 
and off-shell effects tend to quench the Born scattering
processes that involve intermediate hadronic states.

For the considered nucleon model, we find
that scattering from the quark constituent can be significant 
at modest Q values such as Q $>$ 1 to 2 GeV/c.  Once partonic 
scattering becomes dominant, it 
is expected to remain dominant for higher Q.  
Where the transition to dominance of the partonic interactions 
actually takes 
place is a matter of great interest.  The model calculation of this paper 
suggests that this is determined by the size of contact-like contributions,
or equivalently, by the size of the off-forward parton distributions.   
It may occur in some processes at momentum transfer as low as 1 GeV/c
and seems to be likely by 2 GeV/c for the considered isoscalar
meson-exchange current.  

\vspace{0.2in}
Support for this work from the U.S. Department of Energy is
gratefully acknowledged through     
DOE grant DE-FG02-93ER-40762 at the University of Maryland and 
DOE contract DE-AC05-84ER40150, under which the 
Southeastern Universities Research Association (SURA) operates
the Thomas Jefferson National Accelerator Facility.
S. J. W. gratefully acknowledges support of SURA under its Sabbatical
Fellowship Program.  

\appendix
\section{Self-energy and loop integrals} 

Details that have gone into calculations but are 
omitted from the text are collected in this appendix.

The fermion-boson self energy graph, defined in 
Eq.~(\ref{eq:SigmaBubble}), vertex functions defined 
in Eqs.~(\ref{eq:LambdamuLoop}) and (\ref{eq:Lambda5Loop}) 
and the contact-like terms defined in
Eqs.~(\ref{eq:Cmu5loop}) and (\ref{eq:C5muloop}) 
require evaluations of Feynman integrals and 
subsequent reductions of the Dirac matrices to standard forms.  
Integrations over loop momentum k are performed 
by standard methods: n propagator 
factors are combined by means of integrals 
over Feynman parameters $\alpha_1, \alpha_2, 
\cdots, \alpha_{n-1}$, 
into a single denominator function of the
form $[(k - \ell)^2 - F + i \eta]^n$, where the shift 
vector $\ell^{\mu}$ and the function F depend 
upon the external momenta and Feynman parameters.  
Numerator functions involve one 
power of the loop momentum, $k^{\mu}$ for each fermion propagator.  

 Two divergent k-integrations arise and these 
are evaluated by using subtractions. 
The required formulas are,
\begin{eqnarray}
ig\int \frac{d^4k}{(2\pi)^4} \left(\frac{1}{[(k - \ell)^2 - F_{\mu} ]^2} - 
\frac{1}{[(k - \ell)^2 - F_{\Lambda_1} ]^2} \right)   
\Big\{ 1, k^{\mu} \Big\}
 = \frac{g} {16 \pi^2} \ln\left( \frac{F_{\mu}}{F_{\Lambda_1}} \right) 
\Big\{ 1, \ell^{\mu} \big\} ,
\end{eqnarray}
\begin{eqnarray}
2 ig\int \frac{d^4k}{(2\pi)^4} \left(\frac{1}{[(k - \ell)^2 - F_{\mu} ]^3} -
\frac{1}{[(k - \ell)^2 - F_{\Lambda_1} ]^3} \right)  k^{\mu} k^{\nu}  =
\nonumber \\ 
  \frac{ g} {16 \pi^2} \ln\left( \frac{F_{\mu}}{F_{\Lambda_1}} \right) 
\left(-{1 \over 2} g^{\mu \nu} \right) 
+ \frac{ g} {16 \pi^2} \left( \frac{1}{F_{\mu}}- 
\frac{1}{F_{\Lambda_1}} \right)  \ell^{\mu} \ell^{\nu} .
\end{eqnarray}
In all other cases, the k-integrations can 
be performed before subtractions by using
the formulas ( for $n \ge 3$),
\begin{eqnarray}
 \Bigr\{ <1>, <k^{\mu}> , <k^{\mu} k^{\nu}>, 
<k^{\mu}k^{\nu}k^{\sigma}> \Bigr\} \equiv  
\nonumber \\ ig(n-1)!
\times \int \frac{d^4k}{(2\pi)^4} \frac{1}{[(k - \ell)^2 - F + i \eta]^n}
 \Bigr\{ 1, k^{\mu} , k^{\mu} k^{\nu}, k^{\mu}k^{\nu}k^{\sigma} \Bigr\}  =
\nonumber \\
\frac{ (n-3)!g } { (-1)^{(n+1)}16 \pi^2} \Biggr\{ \frac{1}{F^{n-2}}, 
\frac{\ell^{\mu}}{F^{n-2}}, \frac{\ell^{\mu} \ell^{\nu}}{F^{n-2}}
 - \frac{g^{\mu \nu}}{2(n-3)F^{n-3}} , 
\frac{\ell^{\mu} \ell^{\nu}\ell^{\sigma}}{F^{n-2}}
 - \frac{ n^{\mu \nu \sigma }} {2(n-3)F^{n-3}} \Biggr\} 
\end{eqnarray}
where
\begin{equation} 
 n^{\mu \nu \sigma} \equiv  g^{\mu \nu}\ell^{\sigma} 
+ g^{\nu \sigma}\ell^{\mu} + g^{\sigma \mu}\ell^{\nu} .
\end{equation} 

Considering the self energy of an elementary fermion-boson 
bubble graph, we have 
\begin{eqnarray}
 \Sigma_b (p; m, \mu, \Lambda_1)  
   =    i g \int \frac{d^4k}{(2 \pi)^4} \frac{( p\!\!\!/ ~- k\!\!\!/ ~+ m}
{(p-k)^2 - m^2 + i \eta} 
\left(  \frac{1}{k^2 - \mu^2 + i \eta} 
-  \frac{1}{k^2 - \Lambda_1^2 + i \eta} \right).
\end{eqnarray}
Using 
the Feynman parameterization
\begin{equation}
\frac{1}{a b} = \int_0^1 d \alpha \frac{1}{[ \alpha a + (1 - \alpha ) b]^2},
\end{equation}
to combine denominators, we have in this case the 
shift vector $\ell = \alpha p$, and denominator functions
$F_{\mu}$ and $F_{\Lambda_1}$, where the general form is 
\begin{equation}
F_{\Lambda} = \alpha \Lambda^2 + (1 - \alpha) m^2 - \alpha(1 - \alpha) p^2 .
\end{equation}

Integrating over loop momentum produces the two scalar
parts defined in Eq.~(\ref{eq:SigmaAB}), as follows,
\begin{equation}
\left\{ A(p^2), B(p^2) \right\} = \frac{g}{16 \pi^2} 
\int_0^1 d \alpha \ln~\left( \frac{F_{\mu}}
{F_{\Lambda_1}}\right) \left\{ \alpha , m \right\} .
\end{equation}
Using these formulas and the condition $MA(M^2) + B(M^2) = 1$, 
one may determine the coupling constant g 
such that there is a bound state of mass M, where M $<$ m + $\mu$.  
The corresponding formulas for $A'(p^2)$ and $B'(p^2)$ 
are obtained by differentiating with respect to $p^2$,
\begin{eqnarray}
\left\{ A'(p^2), B'(p^2) \right\} = - \frac{g}{16 \pi^2} \int_0^1 d \alpha~ 
\alpha(1 - \alpha) 
\left( \frac{1}{F_{\mu}}
 -  \frac{1}{F_{\Lambda_1}} \right) \left\{ \alpha, m \right\}.
\label{eq:A'B'}
\end{eqnarray} 

   Wave-function renormalization constant $Z_2$ has contributions from the 
elementary bubble graph that may be expressed, 
using Eqs.~(\ref{eq:Zsub2}) and (\ref{eq:A'B'}), as follows,
\begin{eqnarray}
Z_2^{-1} = \frac{ g }{16 \pi^2}  \int_0^1 d \alpha \Biggr\{ 2M
(M\alpha+m)\alpha (1-\alpha) 
\left(
\frac{1}{F_{\mu}} - \frac{1}{F_{\Lambda_1}} \right) 
  -\alpha \ln\Biggr(\frac{F_{\mu}}{F_{\Lambda_1}} 
\Biggr) \Biggr\}
\end{eqnarray} 
and for $Z_2$, $p^2 = M^2$ in $F_{\mu}$ and $F_{\Lambda_1}$. 
Using the identity,  
\begin{eqnarray}
\int_0^1 d \alpha \Biggr\{ 
(1 - \alpha)(m^2 - M^2 \alpha^2) \left( \frac{1}{F_{\mu}} 
- \frac{1}{F_{\Lambda_1}} \right) 
+(2 \alpha -1) \ln \left( \frac{F_{\mu}}{F_{\Lambda_1}} 
\right) \Biggr\} = 0,
\label{eq:identity}
\end{eqnarray}
an equivalent expression for $Z_2^{-1}$ is,
\begin{eqnarray}
Z_2^{-1} &&=  \frac{ g}{16 \pi^2} \int_0^1 d \alpha
  (1-\alpha) \Biggr\{ (M\alpha+m)^2\left( \frac{1}{F_{\mu}}
- \frac{1}{F_{\Lambda_1}} \right) 
-\ln \Biggr( \frac{F_{\mu}}{F_{\Lambda_1}} 
\Biggr) \Biggr\}
\end{eqnarray}
The integral here is the same as for the contribution 
of the elementary bubble graph to 
the normalization of the parton distribution, showing that 
the factor $Z_2$ 
guarantees the normalization as in Eq.~(\ref{eq:fofxnorm}).  
Equation~(\ref{eq:identity}) can be 
verified by integrating by parts the term involving $2 \alpha -1$.

\section{Three-point functions}
 
Three-point functions required for photon and pion vertices are 
defined by Eqs.~(\ref{eq:LambdamuLoop}) and (\ref{eq:Lambda5Loop}).  They
are calculated numerically from formulas involving
integrations over two Feynman parameters, $\alpha_1$ and $\alpha_2$.  The 
denominator function F$_{\Lambda}$  that
results from combining the propagators in the three-point function, assuming 
a generic mass $\Lambda$ for the boson, is
\begin{eqnarray}
F_{\Lambda} =&& \alpha_1 \Lambda^2 + \alpha_2 (m^2 - p_i^2) 
+ (1 - \alpha_1 - \alpha_2)
(m^2 - p_f^2) 
+ \ell^2,
\end{eqnarray}
and the shift vector is
\begin{equation}
\ell =  \alpha_1 p_i + (1 - \alpha_1 - \alpha_2)p_f.
\end{equation}
Moments of loop momentum need to be expanded in terms of the independent
 external momenta, which we choose to be $p_i$ and $p_f$.
For this expansion, we define,
\begin{eqnarray}
<1> = C_0,
\end{eqnarray}
\begin{eqnarray}
< k^{\mu} > = C_{11} p_i^{\mu} + C_{12} p_f^{\mu},
\end{eqnarray}
\begin{eqnarray}
<k^{\mu} k^{\nu}> =&& C_{21} p_i^{\mu}p_i^{\nu} + C_{22} p_f^{\mu}
p_f^{\nu}+C_{23} ( p_i^{\mu} p_f^{\nu} + p_i^{\nu} p_f^{\mu} ) 
+ C_{24} g^{\mu \nu}.
\end{eqnarray}
where the coefficients are calculated from
\begin{eqnarray}
&&\left\{ C_0,  C_{11},  C_{12}, C_{21}, C_{22}, C_{23} \right\} =
\frac{-g}{16 \pi^2} \int_0^1 d \alpha_1 
\int_0^{1-\alpha_1} d \alpha_2 \left( \frac{1}{F_{\mu}} 
-  \frac{1}{F_{\Lambda_1}} \right) \nonumber \\
&& \Bigr\{ 1, (1 - \alpha_1), (1 - \alpha_1 - \alpha_2),
(1 - \alpha_1)^2, (1 - \alpha_1 - \alpha_2)^2  ,
(1 - \alpha_1)(1 - \alpha_1 - \alpha_2) \Bigr\} 
\end{eqnarray}
and the final coefficient is 
\begin{eqnarray}
C_{24} =  \frac{g}{32\pi^2} \int_0^1 d \alpha_1 \int_0^{1-\alpha_1} 
d \alpha_2  \ln  \left( \frac{F_{\mu}}{F_{\Lambda_1}} \right)
\end{eqnarray}
Finally, the Dirac matrices from numerators of 
two fermion propagators are simplified to 
standard forms
with the assistance of projection operators
$L^{\rho_i}(p_i)$ and $L^{\rho_f}(p_f)$.  
Once a factor $p\!\!\!/ _i$ is commuted,
if necessary, to act on $L^{\rho_i}(p_i)$,
it becomes $\rho_i W_i$, where $W_i = \sqrt{p_i^2}$.  
Similarly commuting $p\!\!\!/ _f$ as 
necessary to act on
 $L^{\rho_f}(p_f)$  results in  $\rho_f W_f$.  
The final expressions for form factors are:
\begin{eqnarray}
  F_1&&^{\rho _f,\rho_i} (p_f,p_i)= (\rho_f W_f +m) (\rho_i W_i + m) C_0 
+(\rho_f W_f + m) \rho_i W_i C_{11} \nonumber \\
&&  + \rho_f W_f (\rho_i W_i + m) C_{11} 
+ \rho_i W_i \rho_f W_f C_{21} -  
(p_i - p_f)^2 (C_{22} - C_{23})  
- 2 C_{24}
\end{eqnarray} 
\begin{eqnarray}
 F_2&&^{\rho _f,\rho_i} (p_f,p_i) = -(\rho_f W_f + m) C_{12} 
+ (\rho_i W_i + m) C_{12} - 
\nonumber \\
 && (\rho_i W_i + m) C_{11} - \rho_i W_i C_{21} 
+ (-\rho_f W_f + \rho_i W_i) C_{23}
\nonumber \\
\end{eqnarray} 
\begin{eqnarray}
 F_3&&^{\rho _f,\rho_i}(p_f,p_i) = (\rho_f W_f + m) C_{12} 
+ (\rho_i W_i + m) 
\times(C_{12} - C_{11})
 -\rho_i W_i C_{21} \nonumber \\ &&
 +2 (\rho_f W_f - \rho_i W_i) 
(C_{22} - C_{23}) +(\rho_f W_f + \rho_i W_i) C_{23} 
\end{eqnarray} 
\begin{eqnarray}
 F_5^{\rho _f,\rho_i}(p_f,p_i) = (\rho_f W_f + m) (\rho_i W_i + m) C_{0} 
+ (\rho_f W_f + m) \rho_i W_i C_{11}  \nonumber \\ 
 -(\rho_f W_f + m) (\rho_f W_f + \rho_i W_i) C_{12}
- \rho_i W_i (\rho_i W_i + m) C_{11} \nonumber \\ 
 +(\rho_f W_f + \rho_i W_i) (\rho_i W_i + m) C_{12}  
-  W_i^2 C_{21}- ( W_f^2 -  W_i^2 ) C_{23}
\nonumber \\  - (p_i - p_f)^2 [ C_{22} -  C_{23} ]  
-4 C_{24}
\end{eqnarray} 
Dependences of the form factors on $\rho_i$, $\rho_f$ 
and off-shell momenta are made explicit 
in these formulas.  

\section{Expansion in terms of covariants and matrix elements of Born terms }

It is convenient to expand amplitudes $V^{\mu 5}$ 
and $V^{5 \mu}$ in terms of kinematical covariants
and associated scalar amplitudes.  For the kinematical covariants, we use
helicity matrix elements of a set of eight Dirac operators, as follows,
\begin{equation}
k_1^{\mu} = \bar{u}_{\lambda_f} (p_f)
\gamma _5 \gamma^{\mu} u_{\lambda_i}(p_i)
\end{equation} 
\begin{equation}
k_2^{\mu} = p_i^{\mu} \bar{u}_{\lambda_f} (p_f) 
\gamma _5 q\!\!\!/ u_{\lambda_i}(p_i)
\end{equation} 
\begin{equation}
k_3^{\mu} =  p_f^{\mu} \bar{u}_{\lambda_f} (p_f) 
\gamma _5 q\!\!\!/ u_{\lambda_i}(p_i)
\end{equation} 
\begin{equation}
k_4^{\mu} = q^{\mu} \bar{u}_{\lambda_f} (p_f)
\gamma _5 q\!\!\!/ u_{\lambda_i}(p_i)
\end{equation} 
\begin{equation}
k_5^{\mu} = p_i^{\mu} \bar{u}_{\lambda_f} (p_f) 
\gamma _5 u_{\lambda_i}(p_i)
\end{equation} 
\begin{equation}
k_6^{\mu} = p_f^{\mu} \bar{u}_{\lambda_f} (p_f) 
\gamma _5 u_{\lambda_i}(p_i)
\end{equation} 
\begin{equation}
k_7^{\mu} = q^{\mu} \bar{u}_{\lambda_f} (p_f) 
\gamma _5 u_{\lambda_i}(p_i)
\end{equation} 
\begin{equation}
k_8^{\mu} = \frac{1}{2} \bar{u}_{\lambda_f} (p_f) 
\gamma _5 \left[ \gamma ^{\mu} ,
\gamma ^{\nu} \right]q_{\nu} u_{\lambda_i}(p_i)
\end{equation} 
where $\lambda_i$ and $\lambda_f$ denote the 
helicities of initial and final states. 

The direct Born graph of Eq.~(\ref{eq:Vmu5}) is expanded as follows,
\begin{equation}
V^{\mu,5}_{\lambda_f\lambda_i} = \sum _{n=1}^8 V_{Dn} k_n^{\mu}
\end{equation}
where the scalar coefficients are
\begin{eqnarray}
 V_{D1}^{\rho,+} = \sum_{\rho} F_5^{+,\rho} \frac{1}{D^{\rho}} 
\Big[  (W_{p_i + q} - \rho M)F_1^{\rho,+}  
         + \rho F_2^{\rho,+}((p_i + q)^2 - p_i^2) \Big]
\end{eqnarray} 
\begin{eqnarray}
 V_{D2}^{\rho,+} = \sum_{\rho} F_5^{+,\rho} \frac{1}{D^{\rho}} 
[ -2\rho F_2^{\rho,+} ]
\end{eqnarray} 
\begin{eqnarray}
 V_{D3}^{\rho,+}   = 0 
\end{eqnarray} 
\begin{eqnarray}
 V_{D4}^{\rho,+} = \sum_{\rho} F_5^{+,\rho} \frac{1}{D^{\rho}} 
[ \rho (F_3^{\rho,+} - F_2^{\rho,+} ) ]
\end{eqnarray} 
\begin{eqnarray}
 V_{D5}^{\rho,+} = \sum_{\rho} F_5^{+,\rho} \frac{1}{D^{\rho}} 
[ 2\rho F_1^{\rho,+} ]
\end{eqnarray} 
\begin{eqnarray}
 V_{D6}^{\rho,+}   = 0
\end{eqnarray} 
\begin{eqnarray}
 V_{D7}^{\rho,+} = &&\sum_{\rho} F_5^{+,\rho} \frac{1}{D^{\rho}} 
[ W_{p_i + q} F_3^{\rho,+} + \rho M F_3^{\rho,+} + \rho F_1^{\rho,+}]
\end{eqnarray} 
\begin{eqnarray}
 V_{D8}^{\rho,+} =&& \sum_{\rho} F_5^{+,\rho} \frac{1}{D^{\rho}} 
[ -W_{p_i + q} F_2^{\rho,+} - \rho F_1^{\rho,+} - \rho M F_2^{\rho,+}]
\end{eqnarray} 
In the $V_{Dn}$ expressions, $D^{\rho} 
 = 2 W_{p_i + q} Z_2 [ 1 - B((p_i+q)^2) - \rho W_{p_i+q} A((p_i+q)^2) ]$.

Similarly, the cross Born graph of Eq.~(\ref{eq:V5mu}) is expanded as
follows,
\begin{equation}
V^{5,\mu}_{\lambda_f \lambda_i} = \sum_{\rho} \sum _{n=1}^8 V_{Xn} k_n
\end{equation}
where the scalar coefficients are
\begin{eqnarray}
 V_{X1}^{+,\rho} = \sum_{\rho} \Big[ -(W_{p_f - q} - \rho M ) 
F_1^{+,\rho} + \rho F_2^{+,\rho} ( p_f^2 
- (p_f - q)^2) \Big]  \frac{1}{D^{\rho}} F_5^{\rho,+}
\end{eqnarray} 
\begin{eqnarray}
 V_{X2}^{+,\rho}  = 0 
\end{eqnarray} 
\begin{eqnarray}
 V_{X3}^{+,\rho} = \sum_{\rho}  - 2 \rho F_2^{+,\rho}    
\frac{1}{D^{\rho}} F_5^{\rho,+}
\end{eqnarray} 
\begin{eqnarray}
 V_{X4}^{+,\rho} = \sum_{\rho}  \rho (F_2^{+,\rho} + F3^{+,\rho} ) 
\frac{1}{D^{\rho}} F_5^{\rho,+}
\end{eqnarray} 
\begin{eqnarray}
 V_{X5}^{+,\rho}   = 0 
\end{eqnarray} 
\begin{eqnarray}
 V_{X6}^{+,\rho} = \sum_{\rho} 2 \rho F_1^{+,\rho} 
\frac{1}{D^{\rho}} F_5^{\rho,+}
\end{eqnarray} 
\begin{eqnarray}
V_{X7}^{+,\rho} =&& \sum_{\rho}\Big[-\rho F_1^{+,\rho} +
(W_{p_f - q}+\rho M ) 
F_3^{+,\rho} \Big]
\frac{1}{D^{\rho}} F_5^{\rho,+} 
\end{eqnarray} 
\begin{eqnarray}
V_{X8}^{+,\rho} =&& \sum_{\rho}\Big[-\rho F_1^{+,\rho} 
-(W_{p_f - q}+\rho M ) 
F_2^{+,\rho}  \Big] \frac{1}{D^{\rho}} F_5^{\rho,+}  
\end{eqnarray} 
In the  $V_{Xn}$ expressions, $D^{\rho} (p)  = 2 W_{p_f -q} Z_2 [ 1 -
B((p_f - q)^2) - \rho W_{p_f - q} A((p_f - q)^2) ]$.

\section{Contact-like terms}

Contact-like terms involve three fermion propagators and Dirac matrices
$\gamma_{\mu}$ and $\gamma_5$.  
For $C_{5,\mu}$, we have
\begin{eqnarray}
C^{5 \mu} =\bar{u}(p_f)\Biggr[ i g\int \frac{d^4k}{(2 \pi)^4} 
S(p_f - k;m) \gamma^5  S(p_i +q - k;m) 
\gamma^{\mu} S(p_i - k;m) \nonumber \\ D(k;\mu,\Lambda_1) 
\Biggr]u(p_i).
\end{eqnarray} 
We denote the numerator of this expression as
\begin{eqnarray}
N^{5 \mu}(k) \equiv  \bar{u}(p_f)\Biggr[ (p\!\!\!/ _f - k\!\!\!/ + m)
 \gamma^5 (p\!\!\!/ _i + q\!\!\!/ - k\!\!\!/ + m) 
\gamma^{\mu}  (p\!\!\!/ _i - k\!\!\!/ + m)\Biggr]u(p_i).
\end{eqnarray}
The Feynman parameterization used is 
\begin{equation}
\frac{1}{d_1 d_2 d_3 d_4} = 3!~\int [d\alpha] \frac{1}
{[ \alpha_1 d_1 +\alpha_2 d_2 +\alpha_3 d_3 +\alpha_4 d_4]^4},
\end{equation}
where 
\begin{equation} 
\int [d\alpha] \equiv \int_0^1 d \alpha_1 \int_0^{1 - \alpha_1} 
d \alpha_2 \int_0 ^{1 - \alpha_1 - \alpha_2}
d \alpha_3 ,
\end{equation} 
and
\begin{equation}
\alpha_4 = 1 - \alpha_1 - \alpha_2 - \alpha_3 . 
\end{equation}
This leads to a shift vector
\begin{equation}
\ell = \alpha_4 p_f + \alpha_3 (p_i + q)+ \alpha_2 p_i,
\end{equation}
and a denominator function, for boson mass $\Lambda$,
\begin{eqnarray}
F_{\Lambda} =&& \alpha_4 [ m^2 - p_f^2] 
+ \alpha_3 [m^2 - (p_i + q)^2] + \alpha_2 [m^2 - p_i^2]
+ \alpha_1 \Lambda^2 + \ell^2.
\end{eqnarray}
The required integration is
\begin{eqnarray}
C^{5 \mu} =&& \frac{-6g}{16 \pi^2}  \int [d\alpha] \int \frac{d^4k}{i \pi^2} 
N^{5 \mu} (k) \Biggr( \frac{1}{[ (k-\ell)^2 - F_{\mu}]^4}
- \frac{1}{[ (k-\ell)^2 - F_{\Lambda_1}]^4} \Biggr).
\end{eqnarray}
>From the general rules stated above, one sees that a 
$k^{\mu}$ in the numerator in general is replaced by 
$\ell^{\mu}$ after integration over $k$, but there are 
additional contributions from combinations of 
$k^{\mu} k^{\nu}$ that involve $g^{\mu \nu}$.  
Therefore we write $k^{\mu} = \ell^{\mu} + (k - \ell)^{\mu}$,
and expand in powers of $k-\ell$.   Terms that are odd in $k-\ell$ do not 
contribute because of symmetry.  The parts that do contribute are
\begin{equation}
N^{5 \mu} (k) = N^{5 \mu} (\ell)  + \Delta N^{5 \mu} (k-\ell),
\end{equation}
where 
\begin{eqnarray}
&& \Delta N^{5 \mu} (k) =  \bar{u}(p_f)
\Biggr[ (p\!\!\!/ _f -  \ell\!\!\!/  + m) \gamma^5
 k\!\!\!/ \gamma^{\mu}  k\!\!\!/ +  k\!\!\!/\gamma^5 
(p\!\!\!/ _i - \ell\!\!\!/ + q\!\!\!/  + m) \gamma^{\mu}  k\!\!\!/
+  k\!\!\!/  \gamma^5 k\!\!\!/\gamma^{\mu} (p\!\!\!/ _i  -  \ell\!\!\!/ + m)
\Biggr]u(p_i).
\end{eqnarray}
Integration over k produces
\begin{eqnarray}
C_{5\mu} = \frac{-g}{16 \pi^2} \int [d \alpha ] \Biggr\{  N^{5 \mu} (\ell) 
\left( \frac{1}{F_{\mu}^2} - 
 \frac{1}{F_{\Lambda_1}^2} \right) - {1 \over 2} \langle 
 \Delta N^{5 \mu}\rangle
\times \left( \frac{1}{F_{\mu}} - \frac{1}{F_{\Lambda_1}} \right) \Biggr\},
\end{eqnarray}
where 
\begin{eqnarray}
\langle  \Delta N^{5 \mu}\rangle =  
\bar{u}(p_f)\Biggr[ (p\!\!\!/ _f -  \ell\!\!\!/ + m) \gamma^5
 \gamma^{\alpha} \gamma^{\mu}  \gamma^{\beta}g_{\alpha \beta}  +  
\gamma^{\alpha} \gamma^5 
\times (p\!\!\!/ _i + q\!\!\!/ - \ell\!\!\!/ + m)
\gamma^{\mu}  \gamma^{\beta} g_{\alpha \beta}
 \nonumber \\
+  \gamma^{\alpha}  \gamma^5  \gamma^{\beta}g_{\alpha \beta} 
(p\!\!\!/ _i - \ell\!\!\!/ + m)  
\Biggr]u(p_i).  
\end{eqnarray}

   The resulting expressions are reduced and expanded 
in terms of scalar amplitudes times kinematical covariants 
defined in Appendix C as follows.
\begin{equation}
C^{5 \mu}_{\lambda_f \lambda_i} = \sum_{n=1}^{8} C_{Dn}k^{\mu} _n,
\end{equation}
where 
\begin{equation}
C_{D1} = \int [d\alpha] \Big[ -A_3 -M A_4 - M A_6
(\alpha_2 + \alpha_3 + \alpha_4) \Big] 
\end{equation}
\begin{equation}
C_{D2} = \int [d\alpha] \Big[ -(\alpha_2 + \alpha_3)(A_5 + \alpha_3 A_7)\Big]
\end{equation}
\begin{equation}
C_{D3} = \int [d\alpha] \Big[ - \alpha_4 (A_5 + \alpha_3 A_7) \Big] 
\end{equation}
\begin{equation}
C_{D4} = \int [d\alpha] \Big[  - \alpha_3( A_5 + \alpha_3 A_7) \Big] 
\end{equation}
\begin{eqnarray}
C_{D5} =&& \int [d\alpha] \Big[ A_1 
+ (\alpha_2 + \alpha_3) (A_2 + A_6 + A_8) + A_4 
+(\alpha_2 + \alpha_3)(\alpha_4 - \alpha_2 - \alpha_3) M A_7 \Big]
\end{eqnarray}
\begin{eqnarray}
C_{D6} =&& \int [d\alpha] \Big[ \alpha_4 (A_2-A_6 + \tilde{A}_8) + \alpha_4
(\alpha_4 - \alpha_2 - \alpha_3) 
\times M A_7 \Big] 
\end{eqnarray}
\begin{eqnarray}
C_{D7} =&& \int [d\alpha] \Big[ A_1 + \alpha_3 (A_2+A_8) + \alpha_3
(\alpha_4 - \alpha_2 - \alpha_3) 
\times M A_7 \Big] 
\end{eqnarray}
\begin{equation}
C_{D8} = \int [d\alpha] \Big[ - A_4 -   \alpha_3 A_6 \Big] 
\end{equation}
and where
\begin{equation}
A_1 = \Big[(M+m)^2 - \ell^2\Big] \left( \frac{1}{F_{\mu}^2} 
-  \frac{1}{F_{\Lambda_1}^2} \right)
+ 2 \left(  \frac{1}{F_{\mu}} -  \frac{1}{F_{\Lambda_1}} \right) 
\end{equation}
\begin{eqnarray}
A_2 = \Big[\ell^2 - (M+m)^2 - 2M(M+m)  \Big]  
\left( \frac{1}{F_{\mu}^2} -  \frac{1}{F_{\Lambda_1}^2} \right)
- 3 \left(  \frac{1}{F_{\mu}} -  \frac{1}{F_{\Lambda_1}} \right) 
\end{eqnarray}
\begin{eqnarray}
A_3 = \Big[(M+m)[ 2\ell^2 - 2\ell \cdot p_{\rm int} -m(M+m)] 
-m\ell^2\Big] 
 \left( \frac{1}{F_{\mu}^2} -  \frac{1}{F_{\Lambda_1}^2} \right)
\nonumber \\ 
 -\Big[ 3M+2m\Big]  \left(  \frac{1}{F_{\mu}} - 
 \frac{1}{F_{\Lambda_1}} \right) 
\end{eqnarray}
\begin{equation}
A_4 = \Big[(M+m)^2 - \ell^2\Big] \left( \frac{1}{F_{\mu}^2} 
-  \frac{1}{F_{\Lambda_1}^2} \right)
\end{equation}
\begin{equation}
A_5 = \Big[2 (M+m) -2M(\alpha_2 + \alpha_4)  \Big] 
\left( \frac{1}{F_{\mu}^2} 
-  \frac{1}{F_{\Lambda_1}^2} \right)
\end{equation}
\begin{eqnarray}
A_6 = \Big[m^2 - M^2 - \ell^2 + 2\ell \cdot p_{\rm int}\Big] \left( 
\frac{1}{F_{\mu}^2} -  \frac{1}{F_{\Lambda_1}^2} \right)
+ 3 \left(  \frac{1}{F_{\mu}} -  \frac{1}{F_{\Lambda_1}} \right) 
\end{eqnarray}
\begin{equation}
A_7 = \Big[ -2M \Big] \left( \frac{1}{F_{\mu}^2} 
-  \frac{1}{F_{\Lambda_1}^2} \right)
\end{equation}
\begin{eqnarray}
A_8 = \Big[2M^2 (\alpha_2 + \alpha_3 + \alpha_4) - 2p_i\cdot (p_i + q)
(\alpha_2 + \alpha_3)  
+ 2p_f\cdot (p_i + q) \alpha_4  \nonumber \\ 
+ 2p_i\cdot q \alpha_3 \Big] 
\left( \frac{1}{F_{\mu}^2} -  \frac{1}{F_{\Lambda_1}^2} \right)
\end{eqnarray}
Here $p_{\rm int} = p_i + q$ is the intermediate momentum, 
and $\ell$, $F_{\mu}$ and $F_{\Lambda_1}$ 
are expressed as appropriate for $C^{5 \mu}$. 

A very similar analysis is carried out for the 
crossed contact-like term, $C^{\mu 5}$.  
\begin{eqnarray}
C^{ \mu 5} = \bar{u}(p_f)\Biggr[ i g\int \frac{d^4k}{(2 \pi)^4} 
S(p_f - k;m) \gamma^{\mu}  
 S(p_f -q - k;m) \gamma^5 \nonumber \\ 
 S(p_i - k;m) D(k;\mu,\Lambda_1) 
\Biggr]u(p_i).
\end{eqnarray} 
We denote the numerator of this expression as
\begin{eqnarray}
N^{\mu 5} \equiv  \bar{u}(p_f)\Biggr[ (p\!\!\!/ _f - k\!\!\!/ + m) 
 \gamma^{\mu} (p\!\!\!/ _f - q\!\!\!/ - k\!\!\!/ + m)\gamma^5  %
\nonumber \\
\times (p\!\!\!/ _i - k\!\!\!/ + m)\Biggr]u(p_i).
\end{eqnarray}
Proceeding as before leads to a shift vector
\begin{equation}
\ell = \alpha_4 p_f + \alpha_3 (p_f - q)+ \alpha_2 p_i,
\end{equation}
and a denominator function, for boson mass $\Lambda$,
\begin{eqnarray}
F_{\Lambda} =&& \alpha_4 [ m^2 - p_f^2] 
+ \alpha_3 [m^2 - (p_f - q)^2] + \alpha_2 [m^2 - p_i^2]
+ \alpha_1 \Lambda^2 + \ell^2.
\end{eqnarray}
Numerator factors that produce nonzero results are 
expressed in a similar way as above, 
\begin{eqnarray}
N^{ \mu 5} (k) = N^{ \mu 5} (\ell)  + \Delta N^{ \mu 5} (k-\ell),
\end{eqnarray}
where 
\begin{eqnarray}
 \Delta N^{ \mu 5} (k) =  \bar{u}(p_f)\Biggr[ 
(p\!\!\!/ _f - \ell\!\!\!/ + m) \gamma^{\mu}
 k\!\!\!/ \gamma^5  k\!\!\!/ +  k\!\!\!/\gamma^{\mu} 
(p\!\!\!/ _f - \ell\!\!\!/ - q\!\!\!/  + m) \gamma^5  k\!\!\!/  \nonumber \\
 +  k\!\!\!/  \gamma^{\mu} k\!\!\!/ \gamma^5(p\!\!\!/ _i - \ell\!\!\!/ + m)
\Biggr]u(p_i).
\end{eqnarray}
Integration over k produces
\begin{eqnarray}
C_{\mu 5} =&& \frac{-g}{16 \pi^2} \int [d \alpha ]
\Biggr\{  N^{ \mu 5} (\ell) \left( \frac{1}{F_{\mu}^2} - 
 \frac{1}{F_{\Lambda_1}^2} \right) - {1 \over 2} \langle 
 \Delta N^{\mu 5}\rangle 
 \left( \frac{1}{F_{\mu}} - \frac{1}{F_{\Lambda_1}} \right) \Biggr\},
\end{eqnarray}
where 
\begin{eqnarray}
\langle  \Delta N^{ \mu 5}\rangle =  
\bar{u}(p_f)\Biggr[ (p\!\!\!/ _f - \ell\!\!\!/  + m) \gamma^{\mu}
 \gamma^{\alpha} \gamma^5  \gamma^{\beta}g_{\alpha \beta}  +  
\gamma^{\alpha} \gamma^{\mu} 
\times (p\!\!\!/ _f - q\!\!\!/ - \ell\!\!\!/ + m)\gamma^5  
\gamma^{\beta} g_{\alpha \beta}
\nonumber \\
+\gamma^{\alpha} \gamma^{\mu} \gamma^{\beta}g_{\alpha \beta} 
(p\!\!\!/ _i - \ell\!\!\!/ + m)  
\Biggr]u(p_i). 
\end{eqnarray}

   The resulting expressions are expanded in terms of scalar 
amplitudes times the kinematical covariants,
\begin{equation}
C^{ \mu 5} = \sum_{n=1}^{8} C_{Xn}k^{\mu} _n,
\end{equation}
where 
\begin{equation}
+ \alpha_3 + \alpha_4) \Big] 
C_{X1} = -C_{D1} 
\end{equation}
\begin{equation}
C_{X2} = \int [d\alpha] \Big[ -\alpha_2 (A_5 + \alpha_3 A_7) \Big]
\end{equation}
\begin{equation}
C_{X3} = \int [d\alpha] \Big[ - (\alpha_3 + \alpha_4) 
(A_5 + \alpha_3 A_7) \Big] 
\end{equation}
\begin{equation}
C_{X4} = -C_{D4}
\end{equation}
\begin{eqnarray}
C_{X5} =&& \int [d\alpha] \Big[ \alpha_2 (A_2-A_6 + \tilde{A}_8) + 
\alpha_2(\alpha_2 - \alpha_3 - \alpha_4) 
M A_7 \Big] 
\end{eqnarray}
\begin{eqnarray}
C_{X6} =&& \int [d\alpha] \Big[ A_1 + (\alpha_3 + \alpha_4) (A_2 + A_6 
+ \tilde{A}_8) + A_4  
+(\alpha_3 + \alpha_4)	(\alpha_2 - \alpha_3 - \alpha_4) M A_7 \Big]
\end{eqnarray}
\begin{eqnarray}
C_{X7} =&& \int [d\alpha] \Big[ -A_1 - \alpha_3 (A_2+\tilde{A}_8) 
- \alpha_3(\alpha_2 - \alpha_3 - \alpha_4) 
M A_7 \Big] 
\end{eqnarray}
\begin{equation}
C_{X8} = C_{D8}
\end{equation}
and where
the functions $A_1$ to $A_7$ take the same {\it form} 
as before, except that $p_{\rm int} = p_f -q$, and the
appropriate $\ell$, $F_{\mu}$ and $F_{\Lambda_1}$ for 
$C^{\mu 5}$ must be used.  Also, equalities such as 
$C_{X1} = -C_{D1}$ mean that $C_{X1}$ takes 
the same {\it form} as $-C_{D1}$, but of course must 
be evaluated with the appropriate $\ell$, and so on.  
Function $\tilde{A}_8$ takes a different form from $A_8$, as follows,
\begin{eqnarray}
\tilde{A}_8 = \Big[2M^2 (\alpha_2 + \alpha_3 + \alpha_4) + 
2p_i\cdot(p_f - q) 
\alpha_2   - 2p_f\cdot (p_f-q) 
(\alpha_3 + \alpha_4)  \nonumber \\ 
- 2p_f\cdot q \alpha_3 \Big] 
\left( \frac{1}{F_{\mu}^2} -  \frac{1}{F_{\Lambda_1}^2} \right).
\nonumber \\ && 
\end{eqnarray}

Calculations have been performed in two ways.  One uses 
the expressions given above and the other 
uses expressions that have been developed by use of the 
symbolic manipulation program  SCHOONSCHIP in order to 
reduce the Dirac matrices to the desired forms and
FORMF to calculate the moments of the one-loop graphs~\cite{v}. 
Two independent computer codes were written and 
checked against one another to verify that the algebra and the numerics
was done correctly.

\newpage
\onecolumn

\begin{figure}
\vbox to 6.5 truein {\vss
\hbox to 6.5 truein {\includegraphics{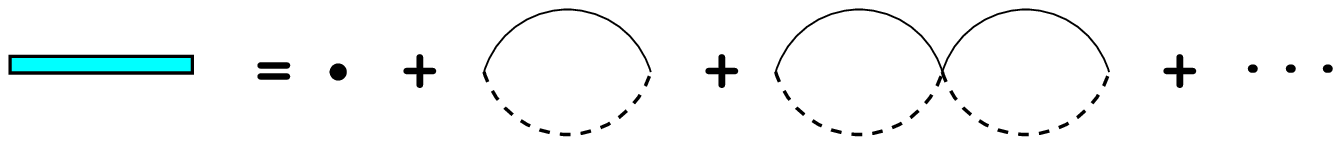}\hss}}
\caption{Bubble graphs that contribute to the composite particle propagator. 
Solid line represents the spin-1/2 quark and dashed line represents 
the spin-0 boson. \label{fig:green}} 
\end{figure}

\begin{figure}
\vbox to 6.5 truein {\vss
\hbox to 6.5 truein {\includegraphics{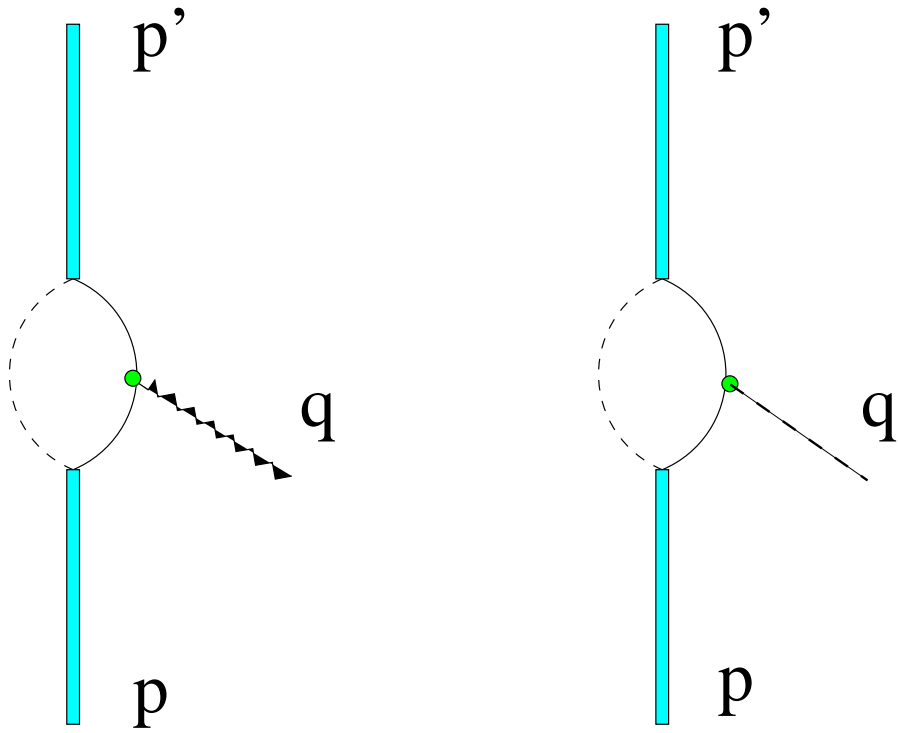}\hss}}
\caption{Photon-quark and pion-quark insertions in the composite 
particle propagator. \label{fig:phvertex}}
\end{figure}

\begin{figure}
\vbox to 7.5 truein {\vss
\hbox to 6.5 truein {\includegraphics{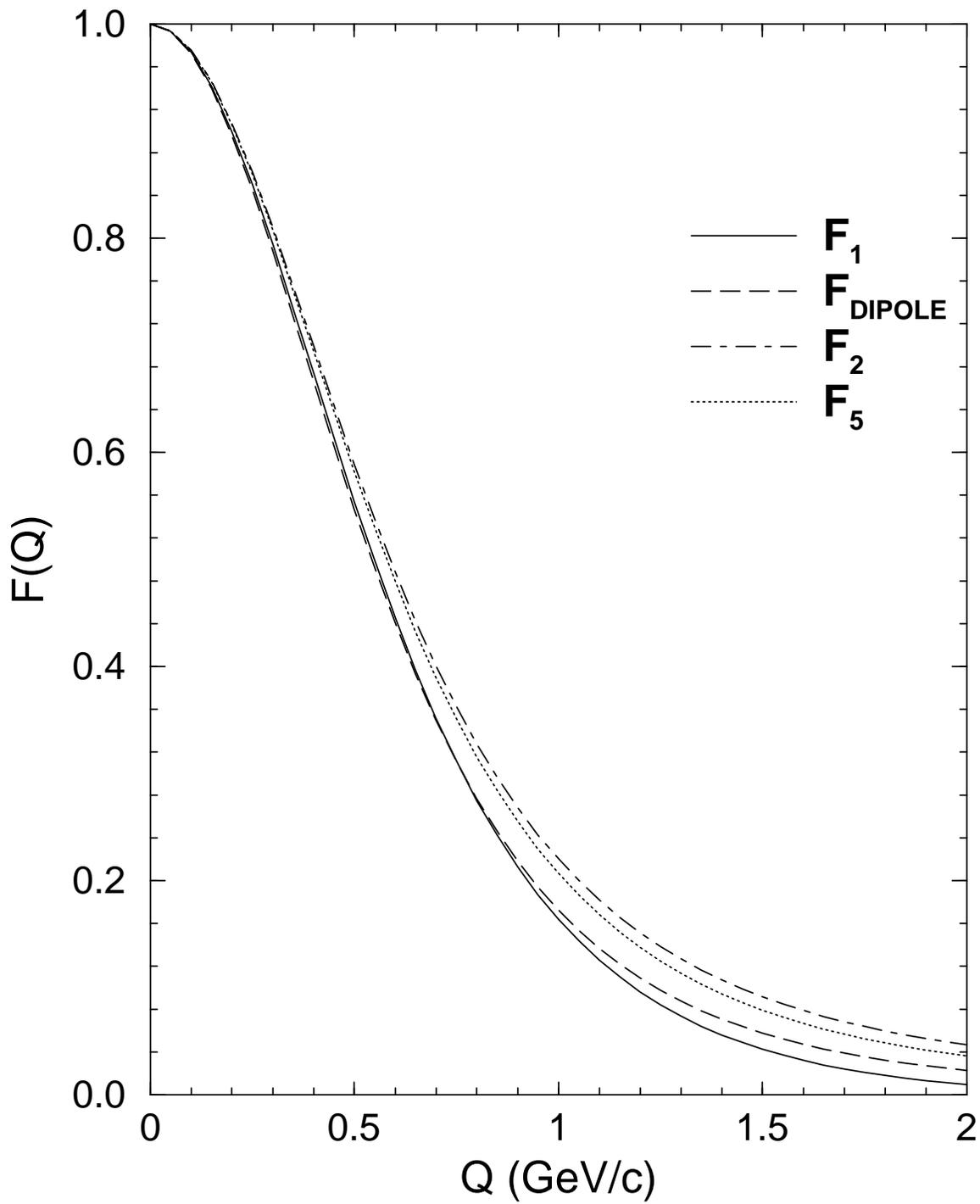}\hss}}
\caption{Electromagnetic and pion form factors for composite particle. 
\label{fig:FormFacs}}
\end{figure}

\begin{figure}
\vbox to 7.5 truein {\vss
\hbox to 6.5 truein {\includegraphics{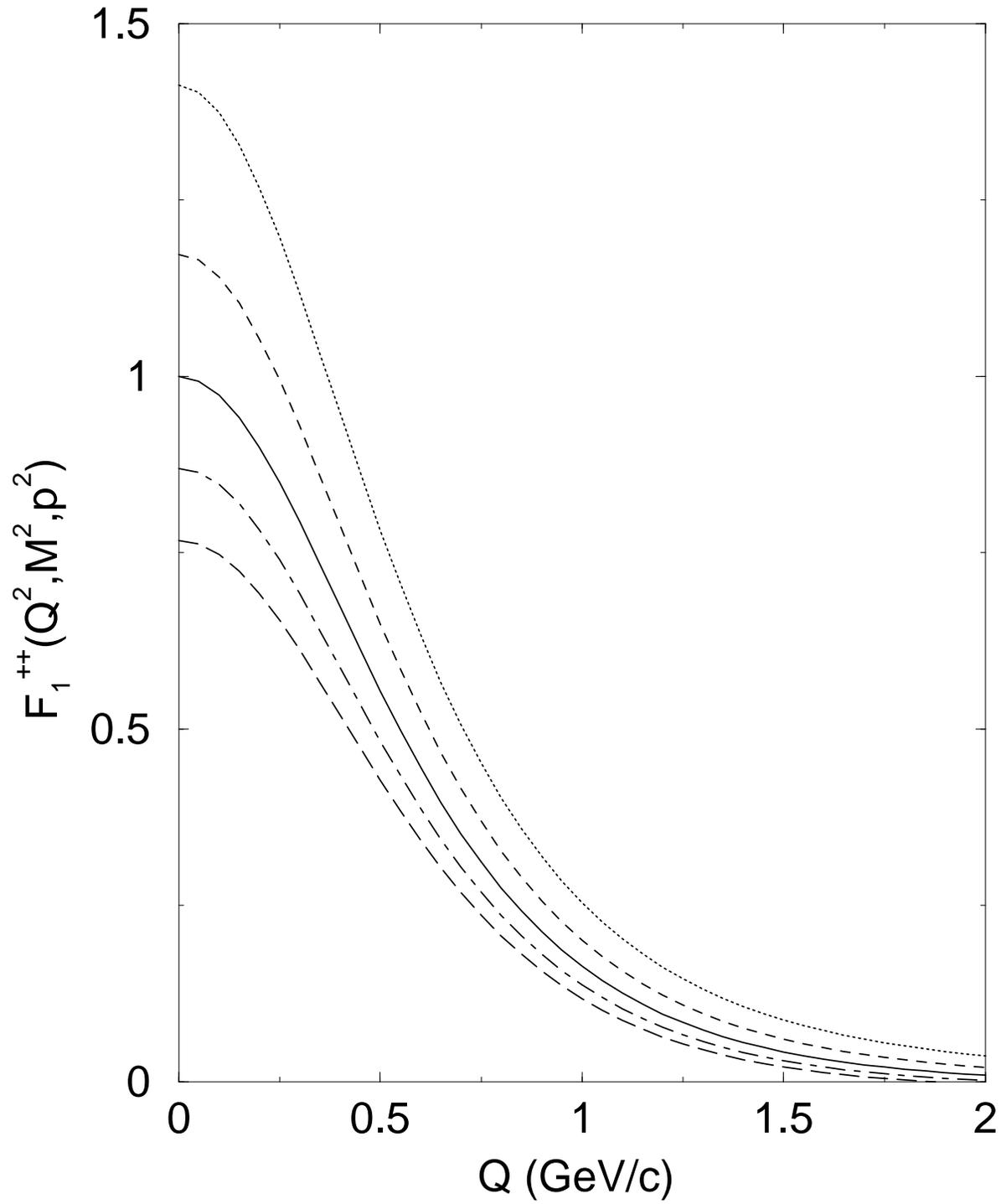}\hss}}
\caption{Dependence of form factor F$_1^{++}$, including propagator 
factor as discussed in text, on off-mass-shell variable 
p$^2$, where p$^2$/M$^2$ = 1.2 (dot line), 1.1 (dashed line), 
1.0 (solid line), 0.9 (dot-dash line and 0.8 (long-dash line).
\label{fig:F1++_off}}
\end{figure}

\begin{figure}
\vbox to 7.5 truein {\vss
\hbox to 6.5 truein {\includegraphics{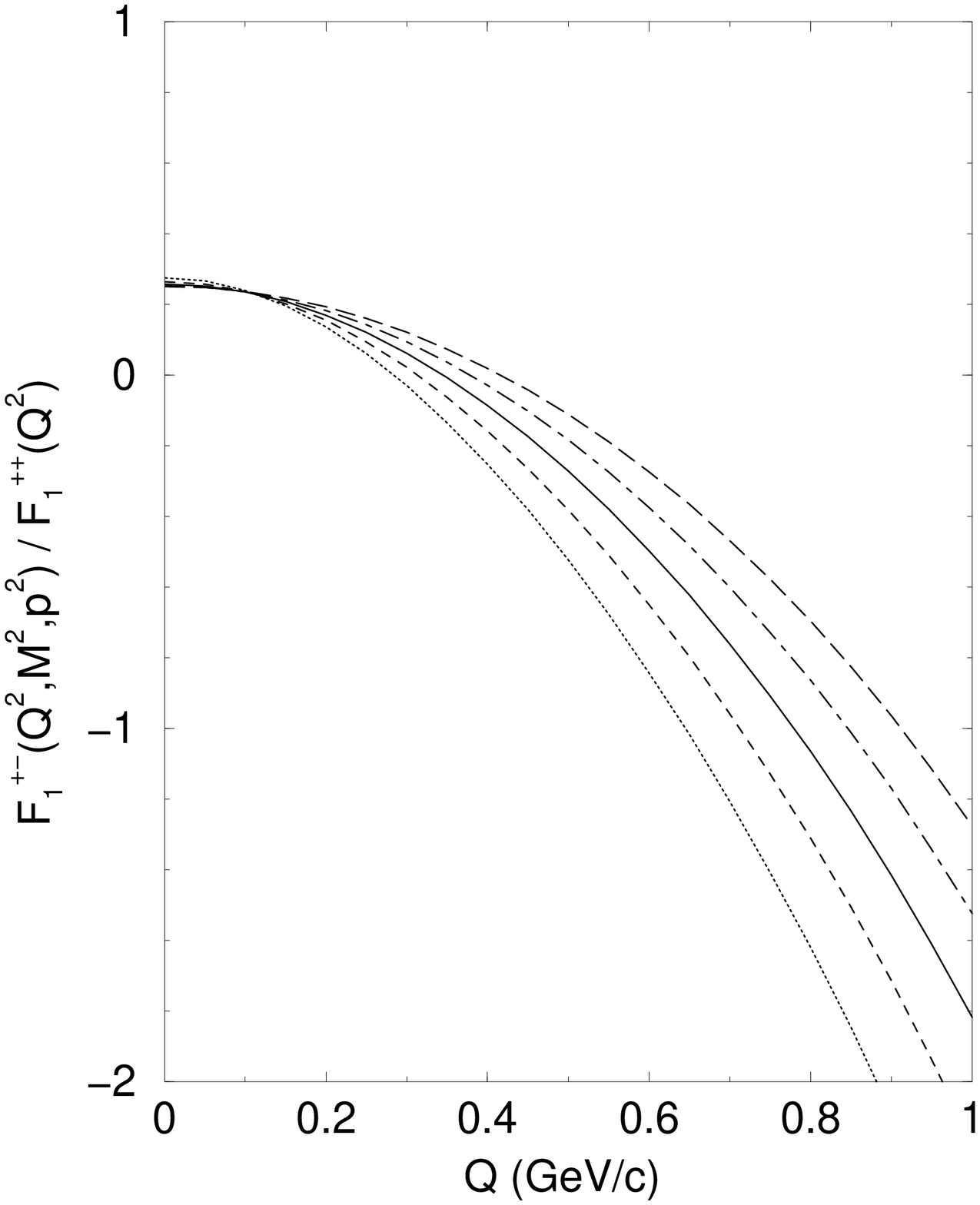}\hss}}
\caption{Dependence of the ratio $F_1^{+-}(Q^2,M^2,p^2)/F_1^{++}(Q^2)$ 
on off-mass-shell variable 
p$^2$, where p$^2$/M$^2$ = 1.2 (dot line), 1.1 (dashed line), 
1.0 (solid line), 0.9 (dot-dash line and 0.8 (long-dash line).
Propagator factor is included as discussed in text.
\label{fig:F1+-_off}}
\end{figure}

\begin{figure}
\vbox to 7.5 truein {\vss
\hbox to 6.5 truein {\includegraphics{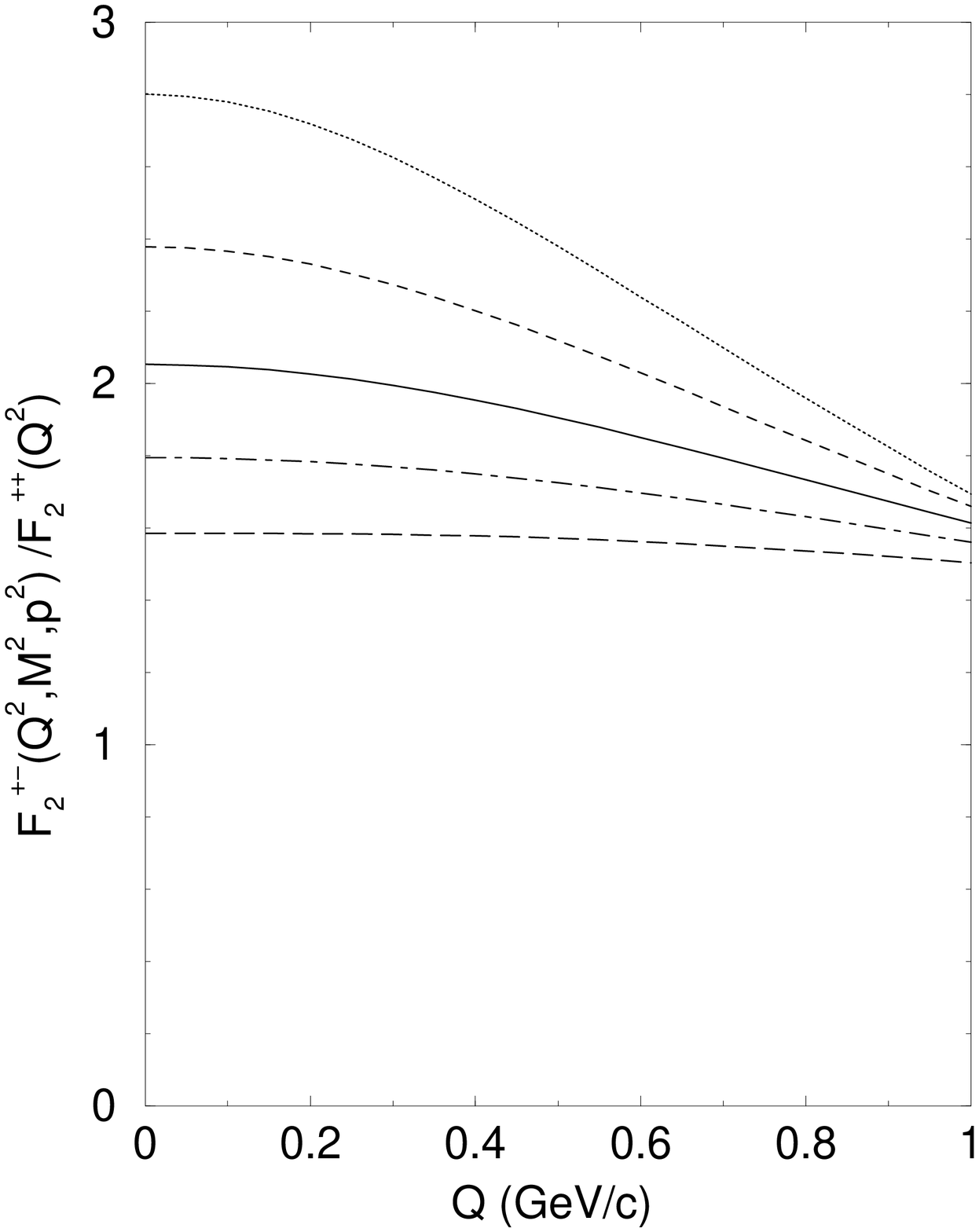}\hss}}
\caption{Dependence of ratio $F_2^{+-}(Q^2,M^2,p^2)/F_2^{++}(Q^2)$ 
on off-mass-shell variable 
p$^2$, where p$^2$/M$^2$ = 1.2 (dot line), 1.1 (dashed line), 
1.0 (solid line), 0.9 (dot-dash line and 0.8 (long-dash line).
Propagator factor is included as discussed in text.
\label{fig:F2+-_off}}
\end{figure}

\begin{figure}
\vbox to 7.5 truein {\vss
\hbox to 6.5 truein {\includegraphics{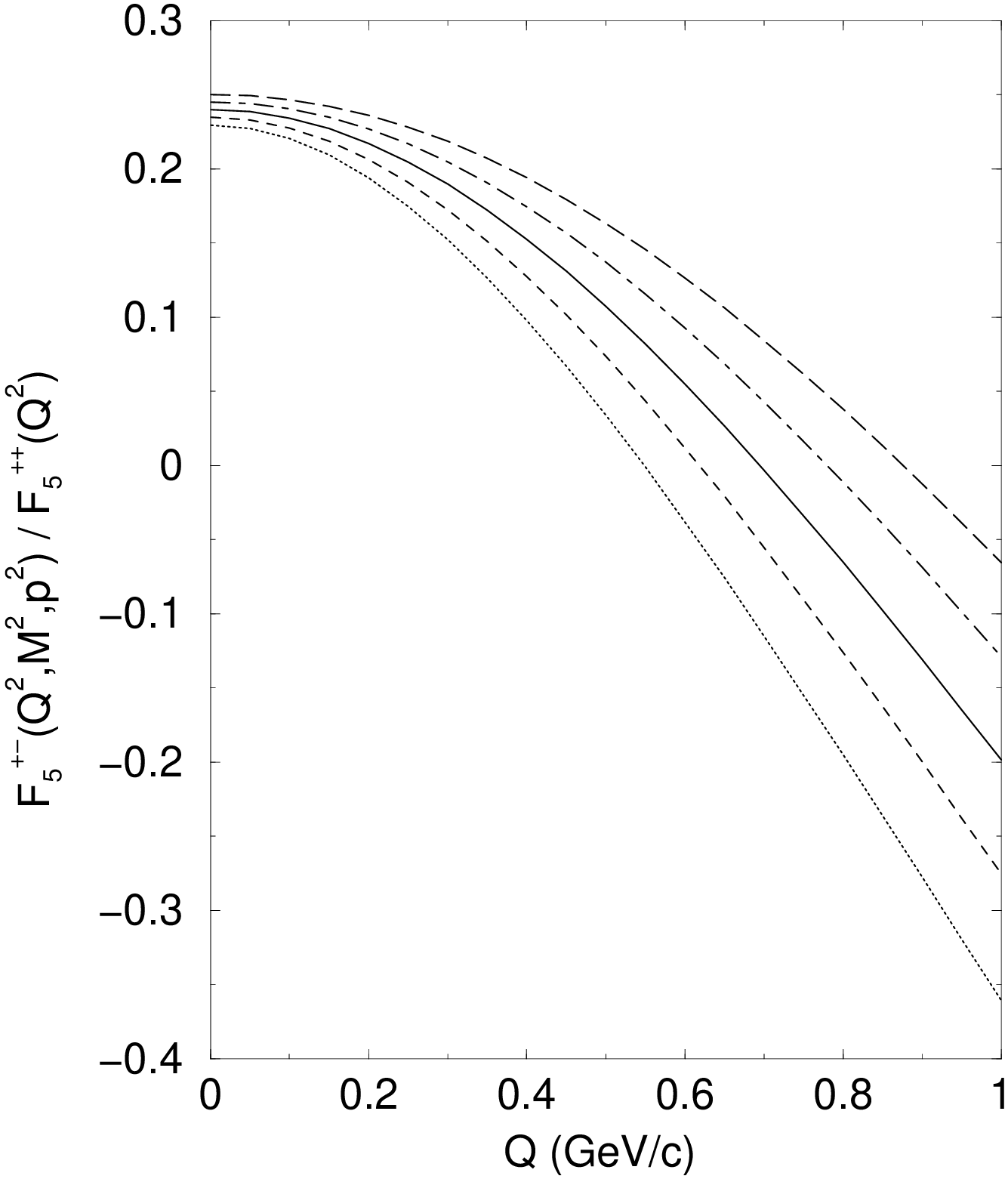}\hss}}
\caption{Dependence of form factor $F_5^{+-}(Q^2,M^2,p^2)/F_5^{++}(Q^2)
$ on off-mass-shell variable 
p$^2$, where p$^2$/M$^2$ = 1.2 (dot line), 1.1 (dashed line), 
1.0 (solid line), 0.9 (dot-dash line and 0.8 (long-dash line).
Propagator factor is included as discussed in text.
\label{fig:F5+-_off}}
\end{figure}

\begin{figure}
\vbox to 7.5 truein {\vss
\hbox to 6.5 truein {\includegraphics{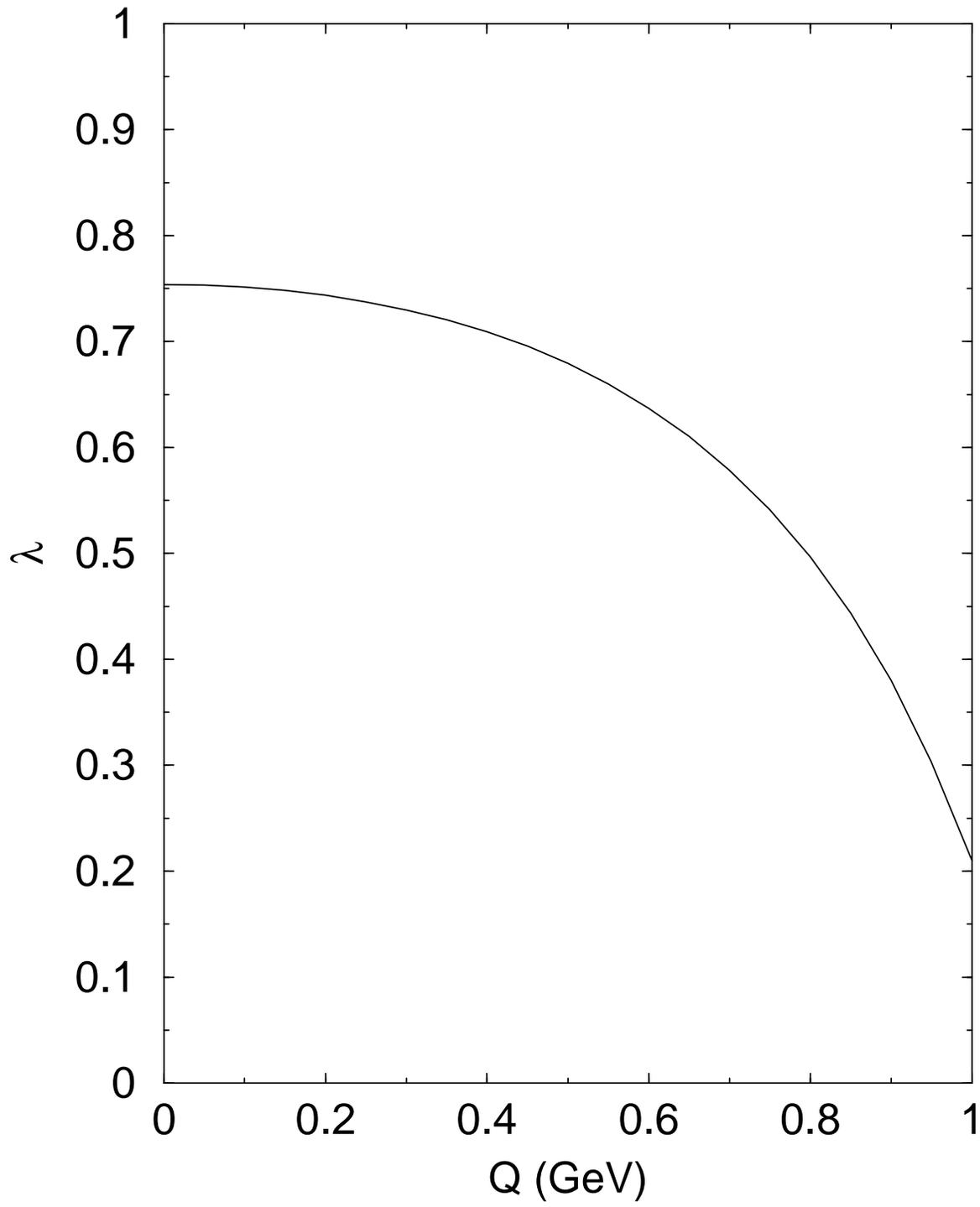}\hss}}
\caption{Pseudovector fraction of pion-nucleon coupling. 
\label{fig:lambda}  }
\end{figure}

\begin{figure}
\vbox to 7.5 truein {\vss
\hbox to 6.5 truein {\includegraphics{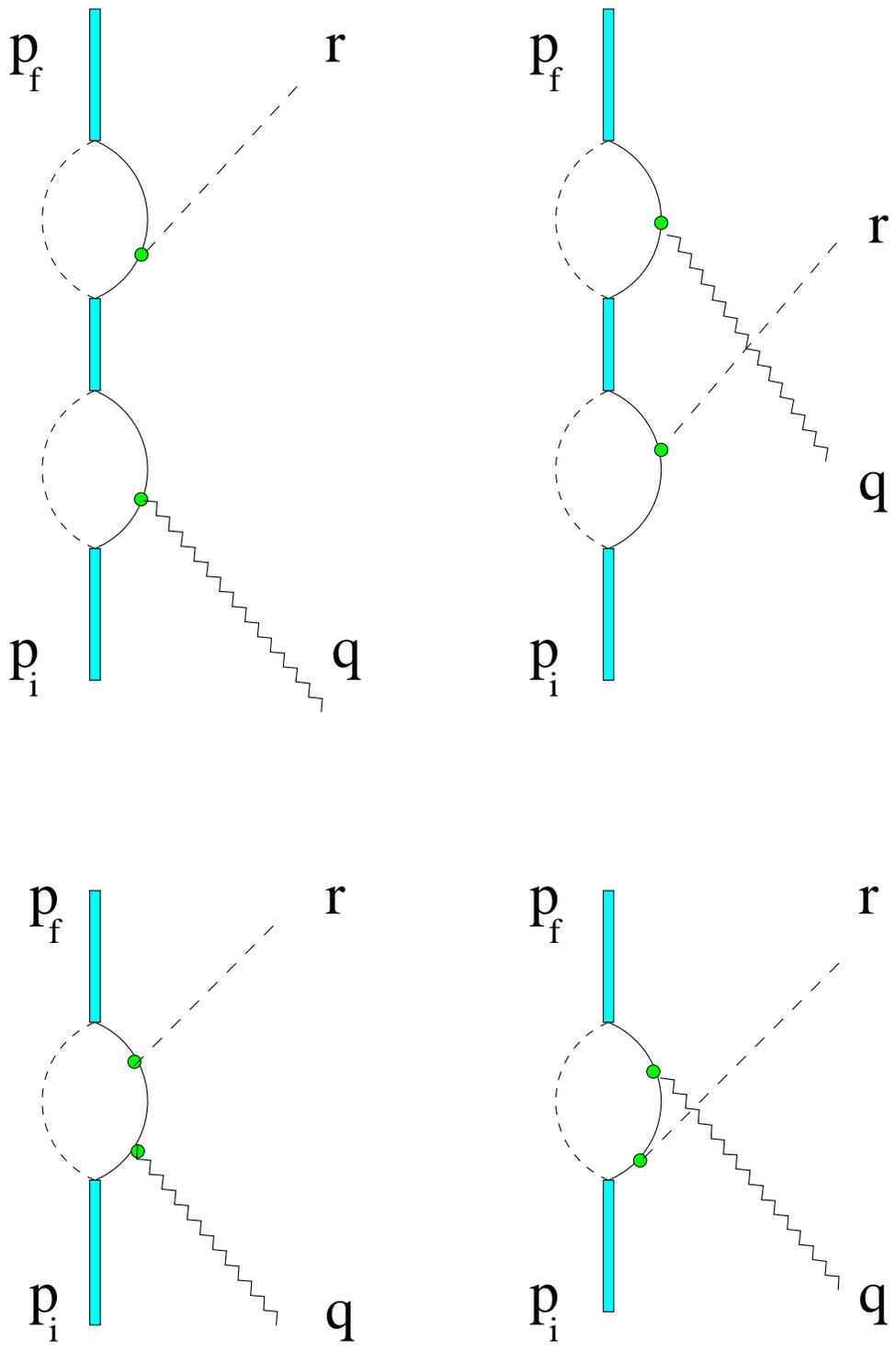}\hss}}
\caption{Photopion amplitudes. Top:  direct and crossed Born graphs.
Bottom: direct and crossed contact-like graphs.
\label{fig:Vmu5}  }
\end{figure}

\begin{figure}
\vbox to 7.5 truein {\vss
\hbox to 6.5 truein {\includegraphics{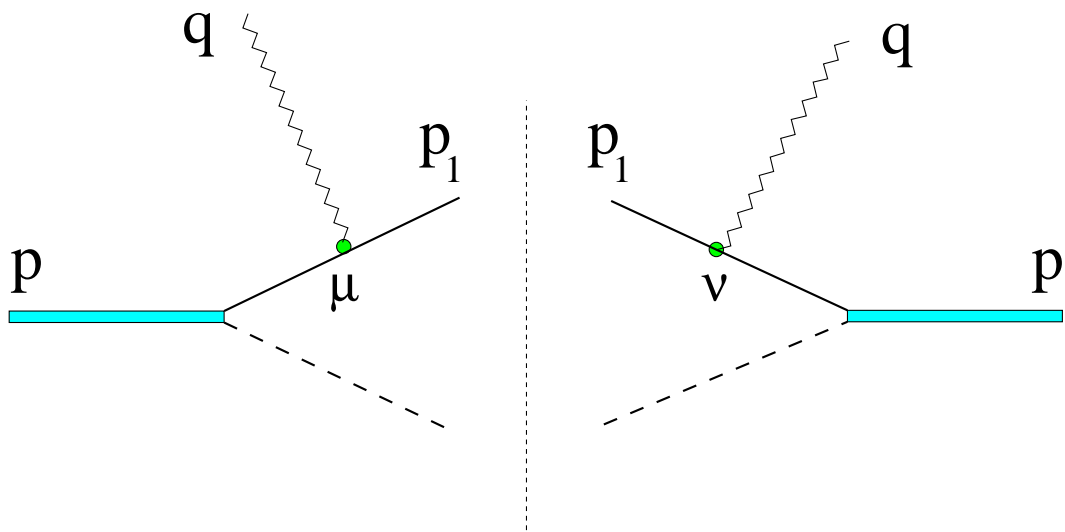}\hss}}
\caption{Impulse approximation for deep inelastic scattering.
\label{fig:distr}}
\end{figure}

\begin{figure}
\vbox to 7.5 truein {\vss
\hbox to 6.5 truein {\includegraphics{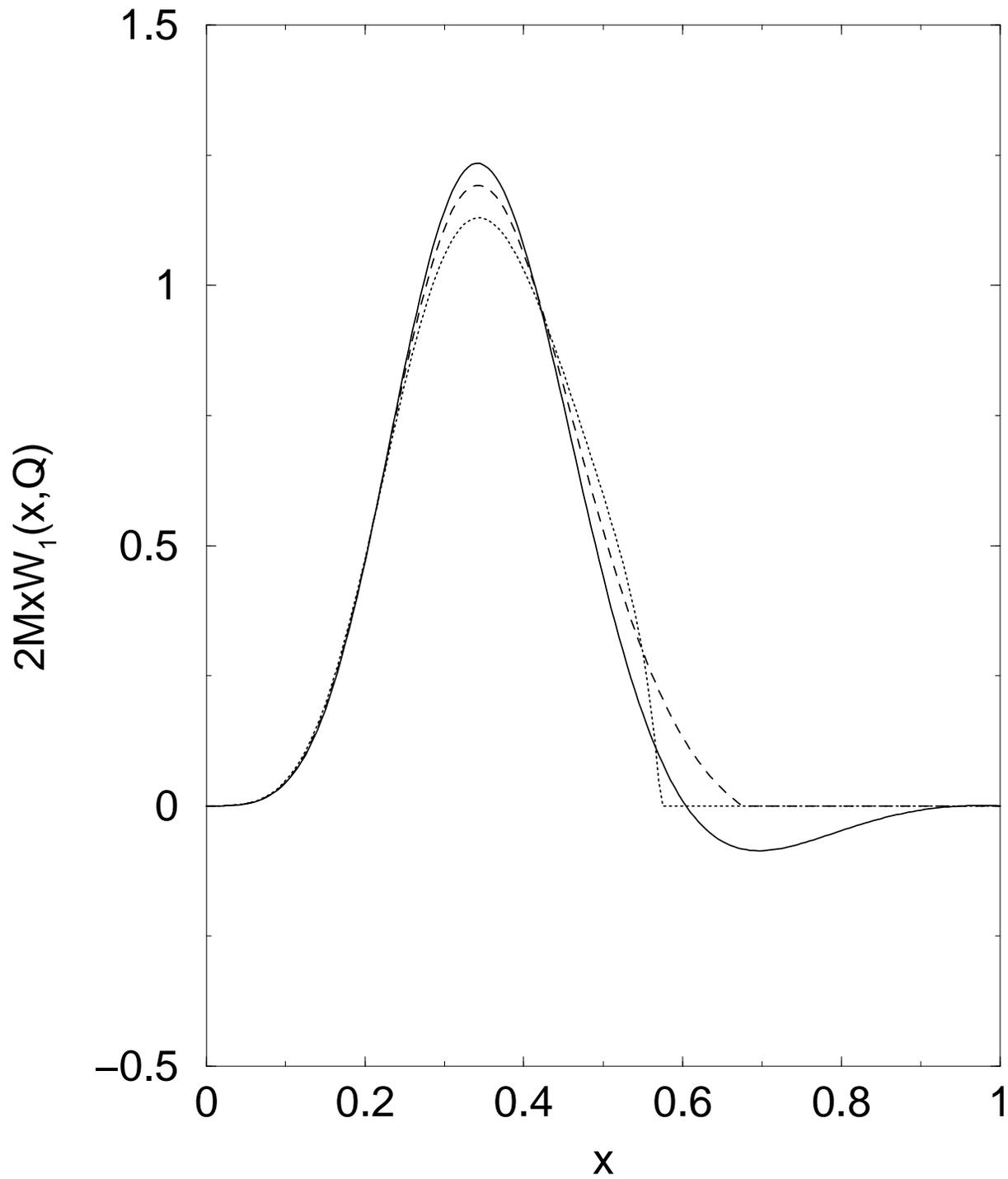}\hss}}
\caption{Structure function for inelastic scattering.
Q = 1.5 GeV/c (dot line), Q= 2.5 GeV/c (dash line), Q = $\infty$ (solid line). 
\label{fig:Wxxplot}}
\end{figure}

\begin{figure}
\vbox to 7.5 truein {\vss
\hbox to 6.5 truein {\includegraphics{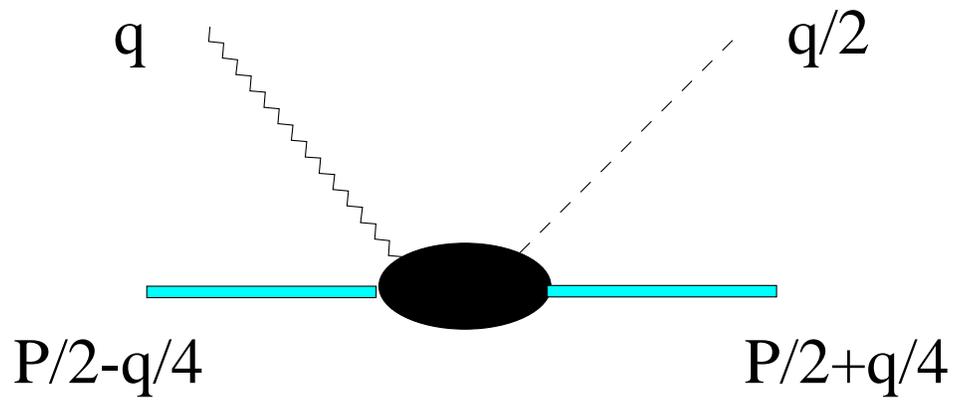}\hss}}
\caption{Quasifree kinematics for photopion amplitude 
in meson-exchange current.  The pion (dashed line) 
is absorbed by a second nucleon, not shown,
such that the each nucleon absorbs half the photon momentum. 
\label{fig:QFkinematics}}
\end{figure}

\begin{figure}
\vbox to 7.5 truein {\vss
\hbox to 6.5 truein {\includegraphics{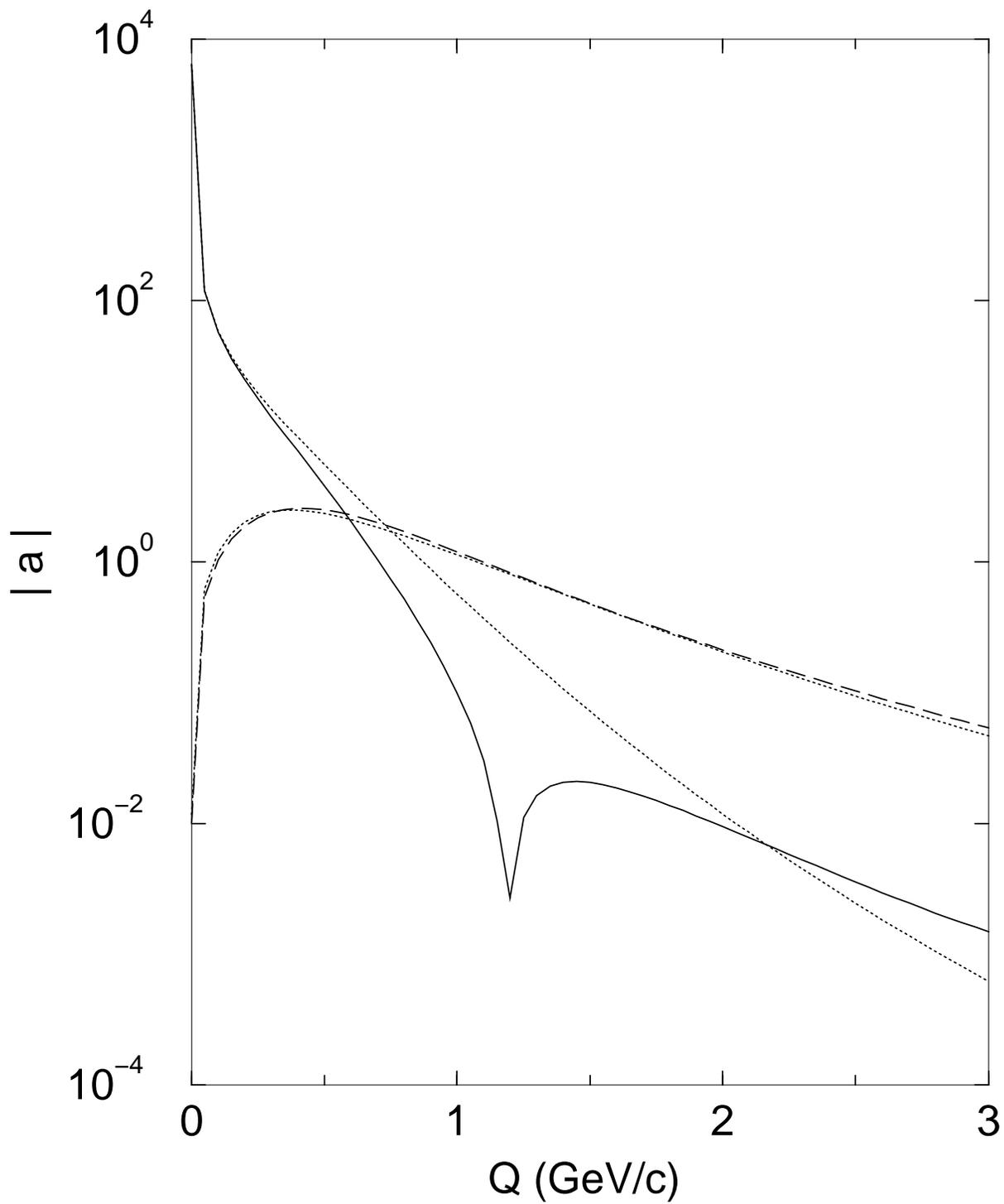}\hss}}
\caption{Born amplitude (solid line) is compared with product of dipole
form factors and propagator (dotted line).  Contact-like contribution
(long dash line) is compared with estimate based on product of dipole form
factor and factor S(Q) = $\kappa^2/(Q^2+ \kappa^2)$, with $\kappa$ = 0.2
GeV$^2$, (dotted line).}  \label{fig:VGVandCEst}
\end{figure}

\begin{figure}
\vbox to 7.5 truein {\vss
\hbox to 6.5 truein {\includegraphics{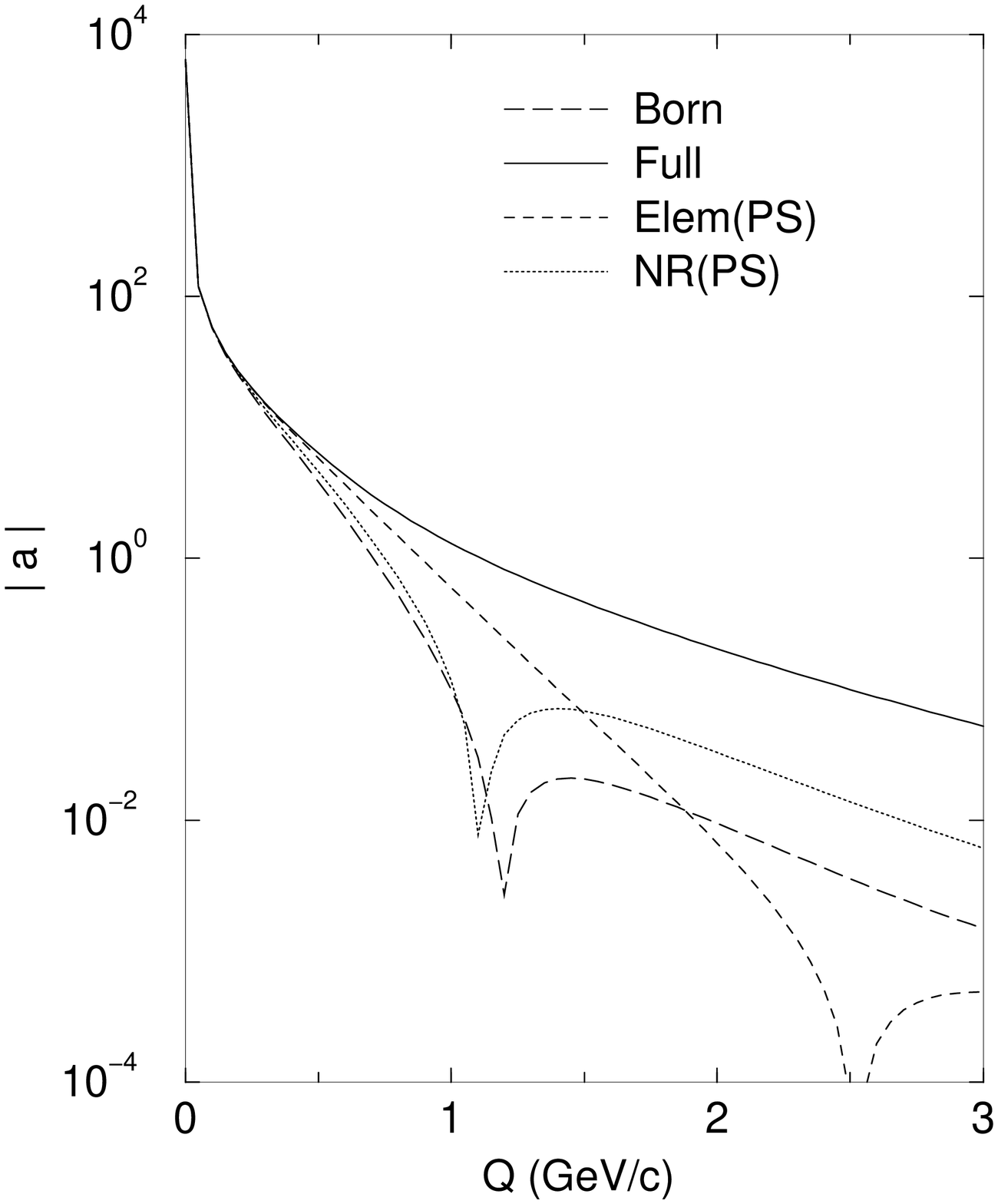}\hss}}
\caption{Isospin-nonflip a amplitude: full amplitude (solid line),
elementary amplitude based on pseudoscalar pion coupling (dash line),
 nonrelativistic amplitude (dot line) and Born amplitude of 
composite model (long dash line). 
\label{fig:aBornFullElem_PS_NR}}
\end{figure}

\begin{figure}
\vbox to 7.5 truein {\vss
\hbox to 6.5 truein {\includegraphics{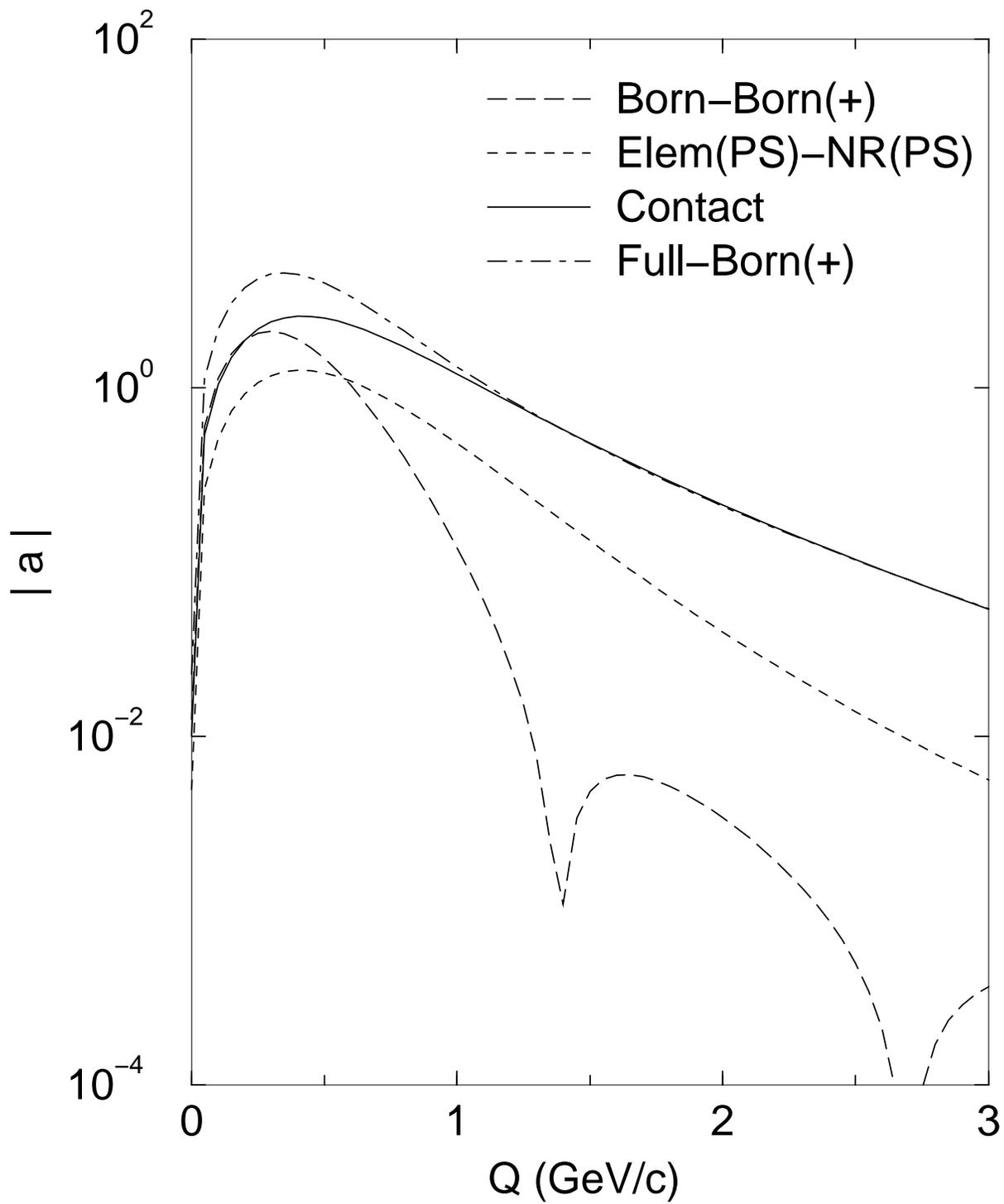}\hss}}
\caption{ Contact-like contribution of composite model (solid line), 
excited states and Z-graph part of Born amplitude for composite 
model (long dash line),
sum of contact-like, excited states and Z-graph parts (dot dash line) and
Z-graph part of pseudoscalar elementary amplitude (dash line).
\label{fig:aElem_PS_ZandCont} }
\end{figure}

\begin{figure}
\vbox to 7.5 truein {\vss
\hbox to 6.5 truein {\includegraphics{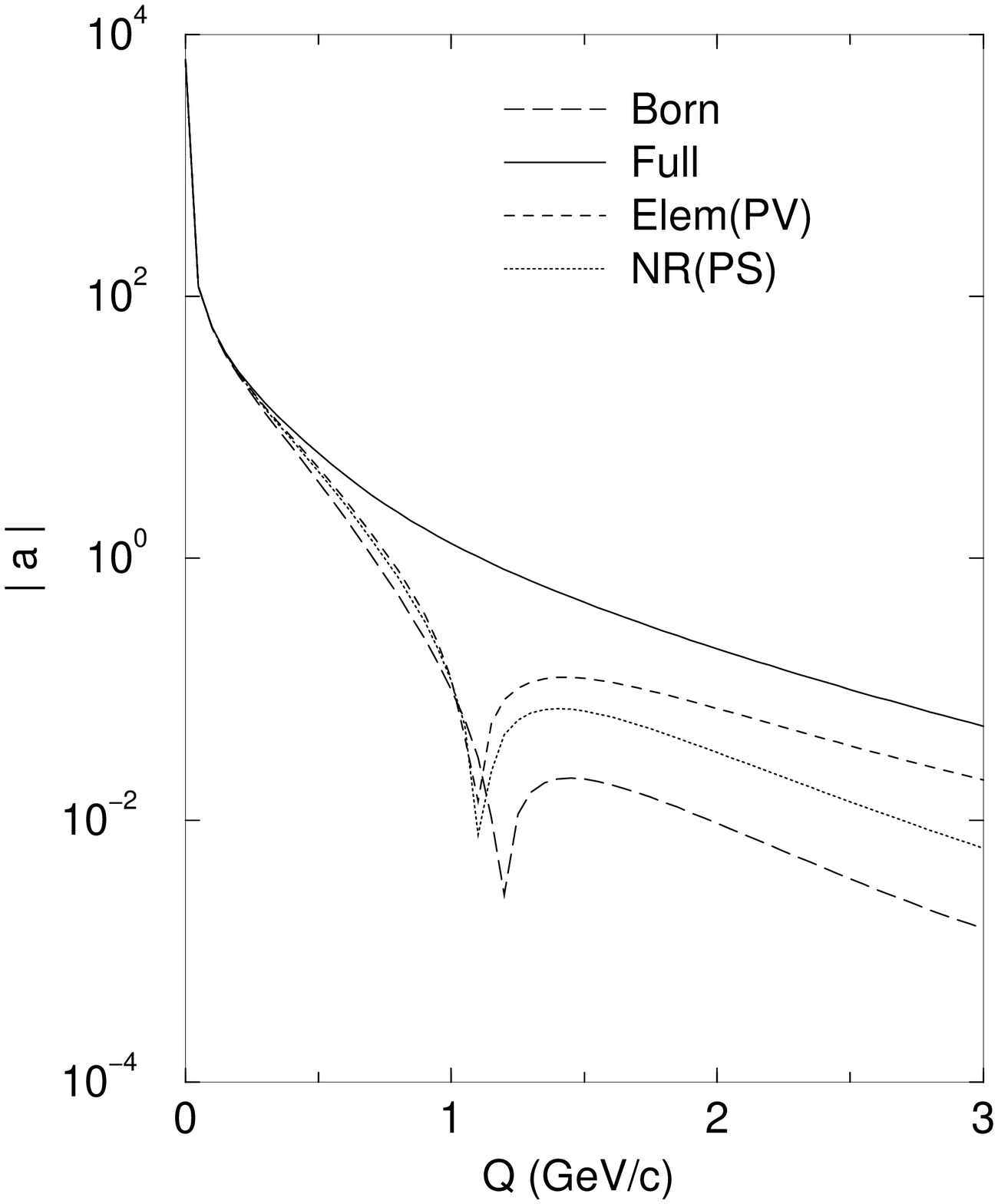}\hss}}
\caption{Isospin-nonflip a amplitude: full amplitude (solid line),
elementary amplitude based on pseudovector pion coupling (dash line),
nonrelativistic amplitude (dot line) and Born amplitude of
composite model (long dash line). 
\label{fig:aBornFullElem_PV_NR}}
\end{figure}

\begin{figure}
\vbox to 7.5 truein {\vss
\hbox to 6.5 truein {\includegraphics{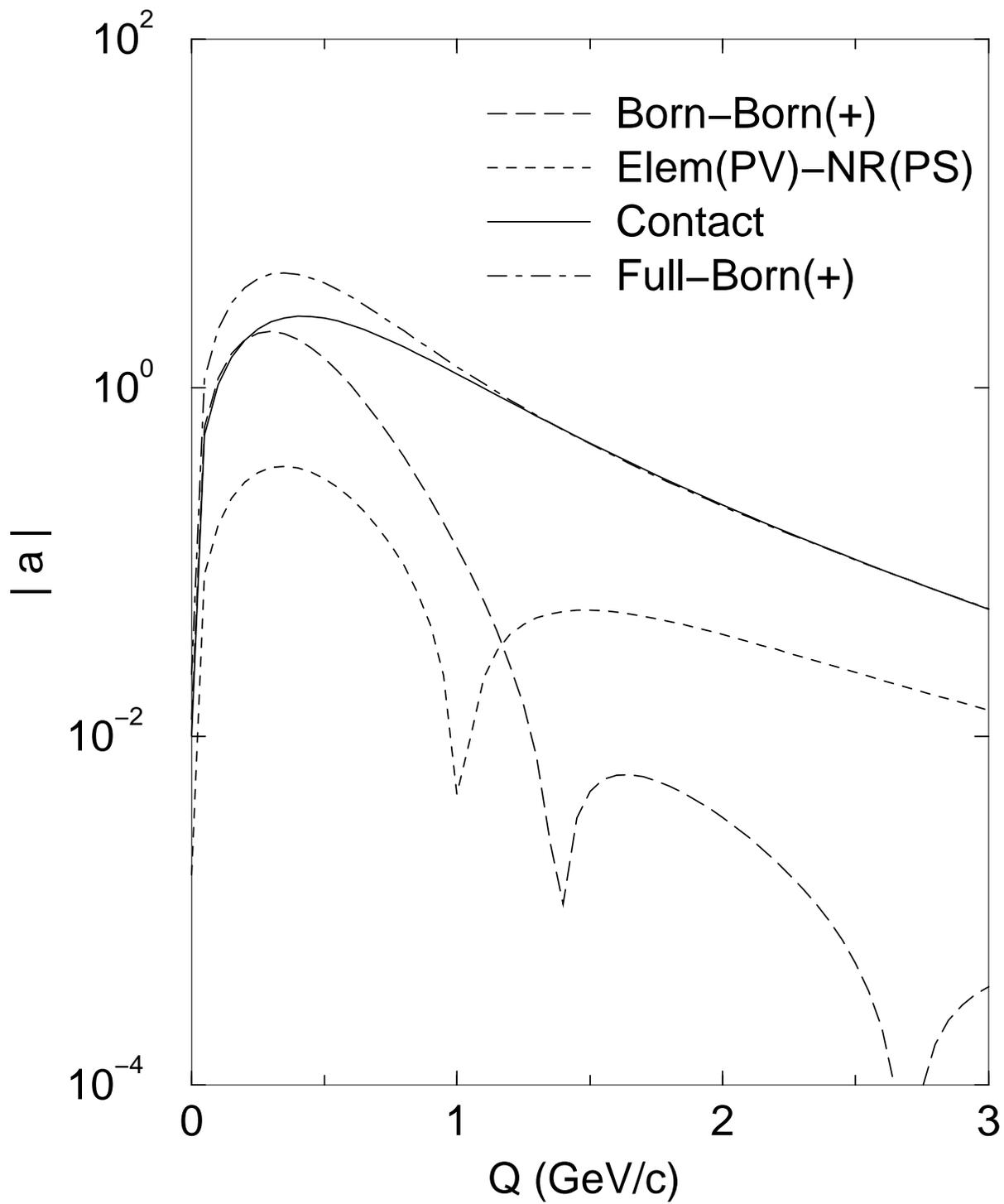}\hss}}
\caption{  Contact-like contribution of composite model (solid line), 
excited states and Z-graph part of Born amplitude for composite 
model (long dash line), sum of contact-like, excited states and Z-graph
parts (dot dash line) and difference between elementary amplitude 
based on pseudovector pion coupling and nonrelativistic 
amplitude (dash line).
\label{fig:aElem_PV_ZandCont} }
\end{figure}

\begin{figure}
\vbox to 7.5 truein {\vss
\hbox to 6.5 truein {\includegraphics{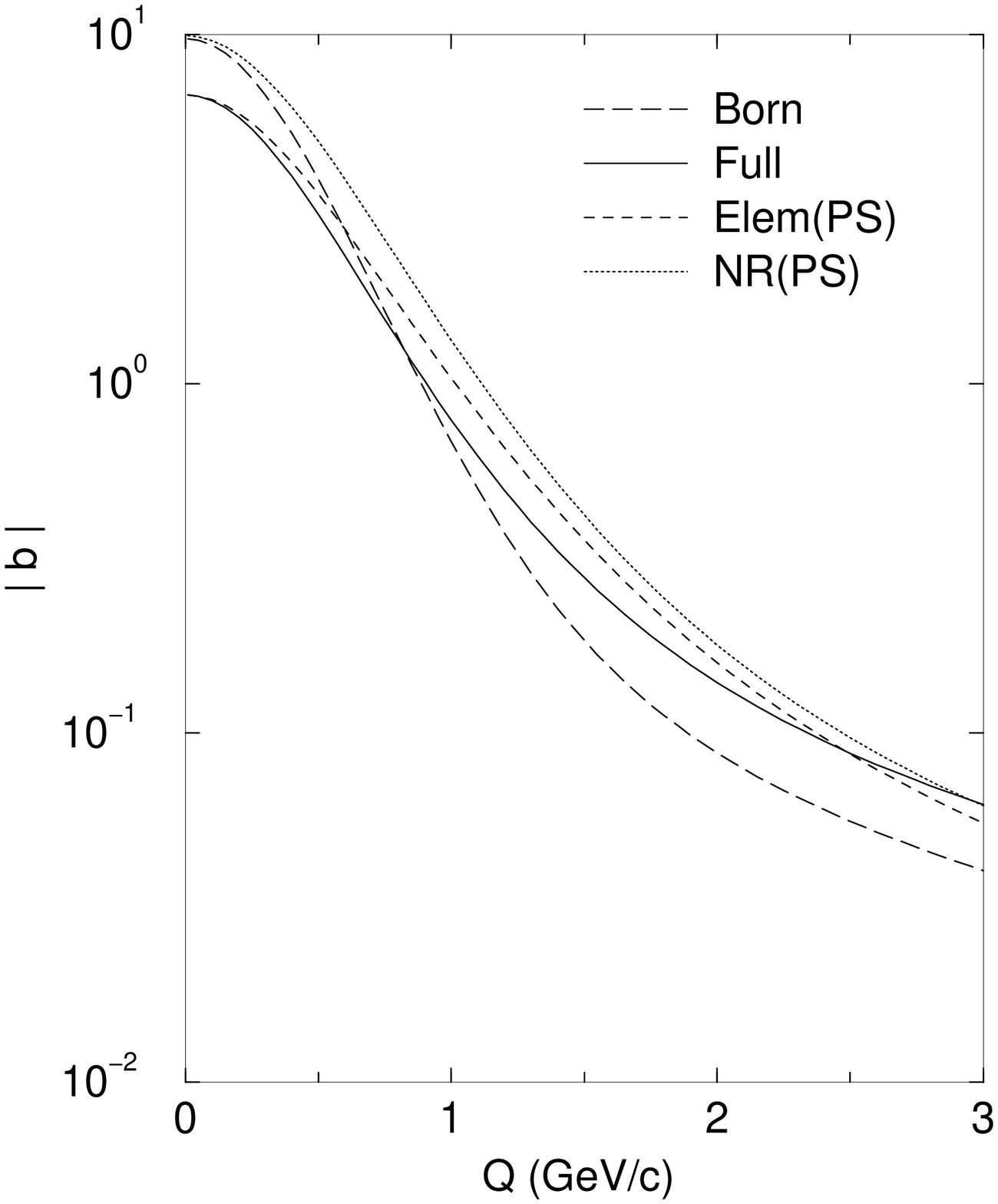}\hss}}
\caption{Isospin-flip b amplitude: full amplitude (solid line),
elementary amplitude based on pseudoscalar
pion coupling (dash line),  nonrelativistic amplitude (dot line) and Born
amplitude of composite model (long dash line). 
\label{fig:bBornFullElem_PS_NR}}
\end{figure}

\begin{figure}
\vbox to 7.5 truein {\vss
\hbox to 6.5 truein {\includegraphics{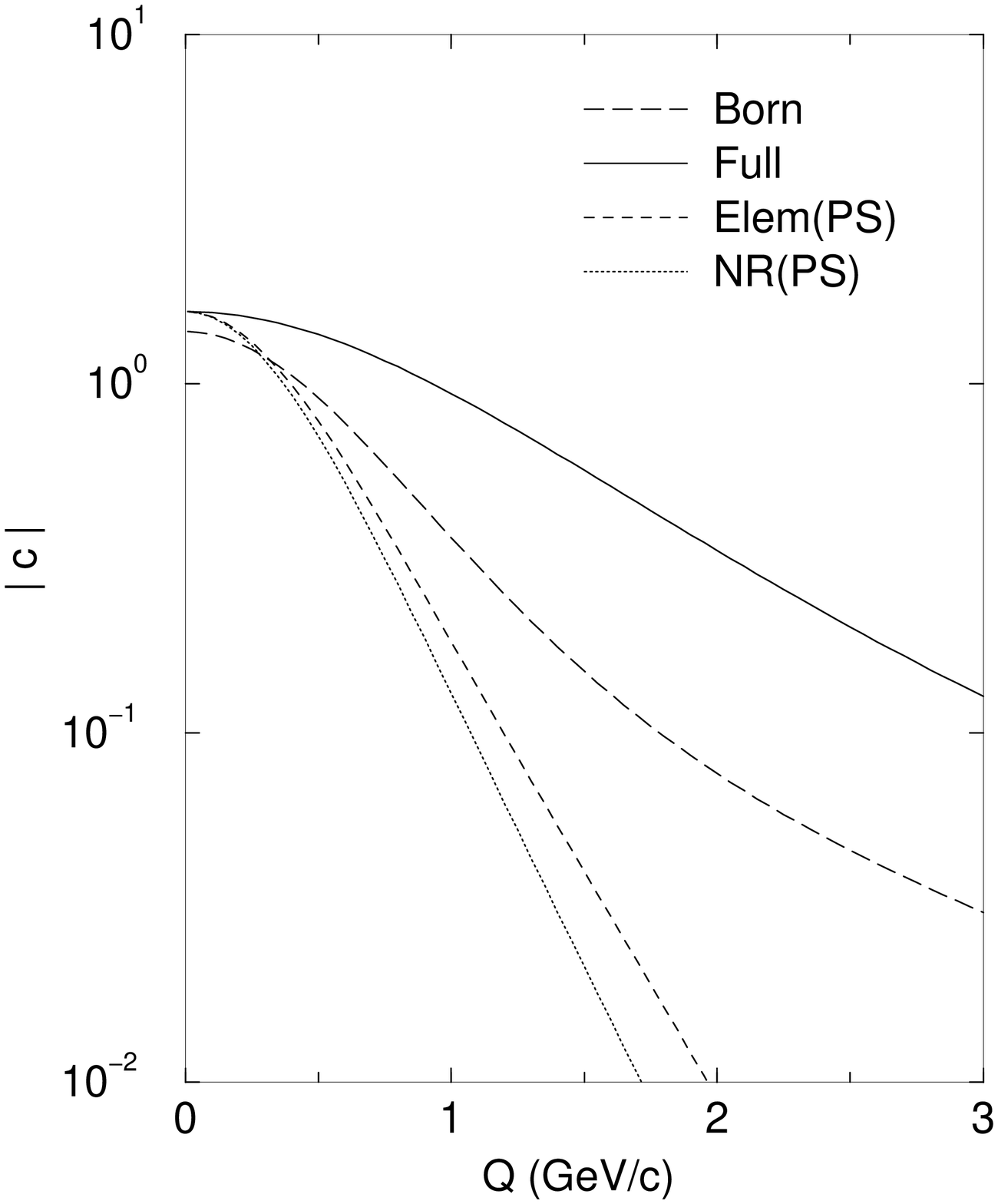}\hss}}
\caption{Isospin-flip c amplitude: full amplitude (solid line),
elementary amplitude based on pseudoscalar
pion coupling (dash line), nonrelativistic amplitude (dot line) and Born
amplitude of composite model (long dash line). 
\label{fig:cBornFullElem_PS_NR}}
\end{figure}

\end{document}